\documentclass[
nohyperref,
reprint,
superscriptaddress,
showkeys,
 amsmath,amssymb,
 aps,
pre,
floatfix,
byrevtex
]{revtex4-1}

\usepackage{graphicx}
\usepackage[english]{babel}
\usepackage{dcolumn}
\usepackage{bm}

\usepackage{float}
\usepackage{placeins}
\usepackage{array}
\usepackage{hyperref}
\hypersetup{
  bookmarksopen = true, 
  bookmarksnumbered = true, 
  colorlinks   = true, 
  urlcolor     = blue, 
  linkcolor    = red, 
  citecolor    = blue 
}

\DeclareUnicodeCharacter{2212}{-}
\usepackage{siunitx}
\DeclareSIUnit\molar{\mole\per\cubic\deci\metre}
\DeclareSIUnit\Molar{\textsc{m}}

\usepackage[version=3]{mhchem} 

\renewcommand{\selectlanguage}[1]{}
\newcolumntype{C}[1]{>{\centering\arraybackslash}p{#1}}
\begin{document}

\preprint{APS/123-QED}

\title{Synthesis, Partial Oxidation and Solvent-dependent Self-Assembly of Ultrathin Cerium Fluoride Nanoplatelets}

\author{Chiara Moretti}
\affiliation
{CNRS, ENS de Lyon, LCH, UMR 5182, 69342, Lyon cedex 07, France}

\author{Damien Alloyeau}
\affiliation
{Université Paris-Cité, CNRS, Laboratoire Matériaux et Phénomènes Quantiques, Paris, 75013 France}

\author{Benjamin Aymoz}
\affiliation
{CNRS, ENS de Lyon, LCH, UMR 5182, 69342, Lyon cedex 07, France}

\author{Laurent Lermusiaux}
\affiliation
{CNRS, ENS de Lyon, LCH, UMR 5182, 69342, Lyon cedex 07, France}

\author{Rodolphe Valleix}
\affiliation
{CNRS, ENS de Lyon, LCH, UMR 5182, 69342, Lyon cedex 07, France}

\author{Benoit Mahler}
\affiliation
{Institut Lumière Matière, Univ Lyon, CNRS, Université Claude Bernard Lyon 1, Villeurbanne, F-69622 France}

\author{Marianne Impéror-Clerc}
\affiliation
{Laboratoire de Physique des Solides, Université Paris-Saclay, CNRS, Orsay, 91405 France}

\author{Benjamin Abécassis}
\email[To whom correspondence should be addressed: ]{benjamin.abecassis@ens-lyon.fr }
\affiliation
{CNRS, ENS de Lyon, LCH, UMR 5182, 69342, Lyon cedex 07, France}
\email{benjamin.abecassis@ens-lyon.fr}

\date{\today}

\begin{abstract}
Two-dimensional colloidal nanoplatelets (NPLs) with atomically defined thickness exhibit unique physical properties, yet understanding their formation mechanism and assembly remains essential for tuning their collective behavior. We report an optimized synthesis of triangular cerium-based NPLs with narrow size and shape distributions \textit{via} thermal decomposition of cerium trifluoroacetate. Combining X-ray diffraction, XPS, and high-resolution STEM, we show that the expected \ce{CeF3} NPL structure undergoes partial oxidation, yielding an oxyfluoride composition \ce{CeOxFy}. Beyond their composition, we investigate how these oleic acid-capped NPLs organize in solution and at interfaces. The choice of solvent governs both the solution-phase organization and the resulting superstructures formed upon evaporation at the liquid–air interface. In solvents that promote face-to-face stacking in solution, evaporation produces films organized into columnar assemblies tens of micrometers long, with the NPL planes oriented perpendicular to the interface. In contrast, solvents in which NPLs remain individually dispersed yield extended hexagonally ordered superlattices with edge-to-edge stacking spanning several micrometers, where the NPLs lie parallel to the interface in an edge-to-edge arrangement. These results highlight that solvent-mediated interactions and pre-existing organization in solution are decisive factors in determining the outcome of evaporative self-assembly of colloidal nanocrystals.
\end{abstract}

\keywords{Cerium fluoride | Ultrathin | Nanoplatelet | Self-assembly | Solvatation forces | SAXS}
\maketitle

\section{Introduction}
Two-dimensional nanomaterials have emerged as a fascinating class of materials that bridge the gap between molecular and bulk systems, offering unique opportunities to explore size- and shape-dependent properties at the nanoscale~\cite{zhang_ultrathin_2015}. Among these, ultrathin colloidal nanoplatelets (NPLs)~\cite{nasilowski_two-dimensional_2016} bear a 2D inorganic core with nanometer-scale thickness while maintaining lateral dimensions on the order of tens of nanometers, and a monolayer of surfactant bound at their surface. In the case of semiconductors, this precise control over thickness leads to strong quantum confinement effects that dramatically impact their optical and electronic properties compared to their bulk counterparts. \cite{diroll_2D_2023} NPLs offer the advantage of being easily synthesized \textit{via} wet-chemical methods with high yields at low cost compared to chemical vapor deposition.

Beyond semiconductors, NPLs made of other materials have remained relatively underexplored. Rare-earth based ultrathin NPL, such as oxides \cite{zhang2016, hudry2015, wang2011, paek2007a, liu2019c, liu2020} or fluorides \cite{stouwdam2002, zhang_SingleCrystalline_2005, du2009,Ye2013}  are promising materials for various applications. Most notably, they can serve as matrices for doping lanthanide-doped nanophosphors, which exhibit unique infrared optical properties \textit{via} efficient upconversion. \cite{sun2022, cheng2022,li2024,luo2025} In this respect, doped fluoride nanocrystals, such as NaYF$_4$, have low phonon energies which reduce non-radiative relaxation processes and thus lead to enhanced luminescence and quantum efficiency compared to oxides. The similarity in ionic radii between O$^{2-}$ and \ce{F-} ions enables the formation of stable mixed oxyfluoride phases \cite{popov1954}, with properties intermediate between those of pure oxide and fluoride compounds \cite{runowski2020}. Exchange between fluoride and oxygen is likely to be easier in ultrathin NPLs, whose large surface-to-volume ratio is expected to enhance interactions with their environment compared to those of three-dimensional nanoparticles. Hence, the propensity of rare-earth fluoride nanoparticles to oxidize is a significant matter that, to our knowledge, has not been previously studied. 

The self-assembly of colloidal nanoparticles is also of paramount importance, as collective properties can emerge from short-range interactions within particle assemblies~\cite{bassani2024, boles2016}. These assemblies are governed by a delicate balance of forces, including van der Waals attractions, steric repulsion from surface ligands, and entropic contributions from both particles and solvent molecules~\cite{widmer-cooper_LigandMediated_2016, monego_colloidal_2018,kister_colloidal_2018, monego_when_2020, schapotschnikow_understanding_2009}. For anisotropic particles like NPLs, the situation becomes even more complex due to the interplay between shape-dependent interactions and orientational degrees of freedom~\cite{Ye2013, liu2021}. Different assembly protocols, such as tuning solvent quality or controlling solvent evaporation, have been optimized to yield a wide diversity of self-assembled 2D and 3D superstructures. The type of structure obtained (crystalline, glassy, or without order) results from an interplay between the interparticle interactions and the destabilization kinetics. It is usually assumed that particles must have time to organize during assembly to form long-range crystalline assemblies and avoid kinetic traps. The choice of solvent plays a particularly critical role, as it affects both the effective inter-particle interactions through solvent-mediated forces and the kinetics of self-assembly during solvent evaporation. In the case of CdSe NPLs \cite{jana_cdse_2016, jana_ligand-induced_2017, guillemeney_Curvature_2022}, stacks are observed in solution~\cite{jana_stacking_2015} before evaporation, and molecular dynamics simulations showed that solvation forces play an important role~\cite{petersen2022, chen2025} in their colloidal stability. Solvent effects are also critical during self-assembly: fast-evaporating alkanes yield face-down assemblies, while those with lower vapor pressures favor edge-up configurations~\cite{gao_cdse_2017, momper_kinetic_2020-1, marino2023}. A full predictive understanding of NPL evaporative self-assembly, therefore, requires characterizing the solution structure prior to evaporation to relate the final structure to potential pre-evaporation assemblies in solution.

In this work, we report a comprehensive investigation of triangular cerium-based NPL synthesized from the thermal decomposition of cerium trifluoroacetate. We developed an optimized synthetic protocol that yields monodisperse triangular NPLs with well-controlled size and shape. We performed detailed structural characterization, revealing their mixed oxyfluoride composition. By systematically varying the dispersing solvent, we demonstrate significant differences in assembly behavior, ranging from well-dispersed individual particles to highly ordered columnar stacks and extended 2D superlattices. Our results reveal clear correlations between solvent properties and assembly outcomes, providing insights into the fundamental mechanisms governing NPL organization and establishing guidelines for controlling their collective arrangement.

\section{Results and discussion}

\subsection{Synthesis of triangular nanoplatelets}

Triangular NPLs were synthesized through the thermal decomposition of cerium trifluoroacetate (\ce{Ce(CF3COO)3} hydrate) in a mixture of oleic acid (OA) and octadecene (ODE)~\cite{zhang_SingleCrystalline_2005}. 
The mixture was heated under vacuum to 100 $^{\circ}$C for 30 minutes, with vigorous magnetic stirring, to remove water and oxygen. The flask was then placed under argon flux, heated to 260 $^{\circ}$C, and held at that temperature for 1 hour. At the end of the reaction, the NPLs were purified several times by centrifugation using a solvent/anti-solvent mixture (see Experimental section). Purification is essential to remove impurities and minimize the amount of free oleic acid that may induce face-to-face stacking~\cite{Ye2013}.

By strictly adhering to the protocol reported in reference~\cite{zhang_SingleCrystalline_2005}, we obtained NPLs with irregular shapes and a large polydispersity, as shown in Fig.~\ref{figSI:TEM_1}a. 
NPL shape control was achieved by tuning the precursor concentration, the ODE/OA ratio, and the reaction temperature. To synthesize NPLs with a triangular shape and a low polydispersity, the experimental conditions were modified by changing the OA and ODE amount (from 20 mmol to 31 mmol), the time under vacuum at 100 $^{\circ}$C (from several minutes to at least half an hour), and the reaction temperature (260 $^{\circ}$C instead of 280 $^{\circ}$C). The increase of OA concentration favored the formation of triangular NPLs with well-defined vertices (Fig.~\ref{figSI:TEM_1}c.) 
Heating at lower temperatures (260 $^{\circ}$C instead of 280 $^{\circ}$C) than in the original protocol favors the formation of homogeneous triangular NPLs with sharp edges, while higher temperatures (300 $^{\circ}$C) yield NPLs with a high polydispersity in size and shape (Fig.~\ref{figSI:TEM_1}b.). Large triangular or hexagonal NPLs with dimensions of 30-70 nm were formed in this case. 

\begin{figure*}
    \centering
    \includegraphics[width=0.9\linewidth]{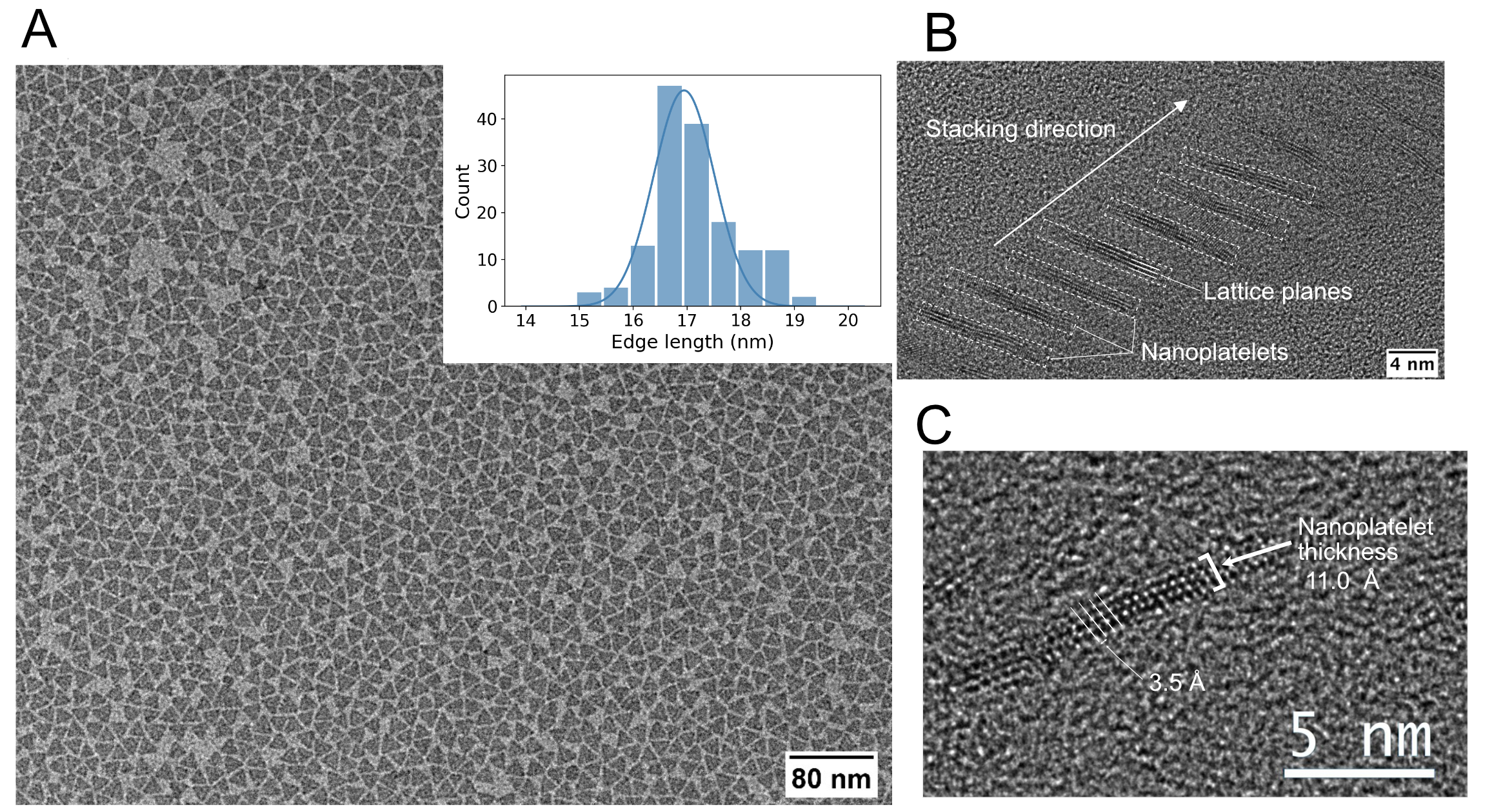} 
    \caption{a) TEM image of purified triangular \ce{CeOxFy} NPLs and histogram of edge length. b) HRTEM image of several triangular NPLs stacked face-to-face (in a column). c) Magnification of a triangular NPL stacked on its edge showing the crystallographic layers.} 
    \label{fig:Fig1METSA}
\end{figure*} 

TEM at low magnification shows the purified product is very homogeneous (Fig.~\ref{fig:Fig1METSA}a), with all nanoparticles being triangular NPLs with an average edge length of 17.0 nm, low polydispersity (relative standard deviation of 5\%), and slightly rounded corners. High resolution STEM (Fig.~\ref{fig:Fig1METSA}b and c of stacked NPL with their edges perpendicular to the substrate proves the NPL ultrathin nature with 3 to 4 atomic layers along their thickness. 

The thermal decomposition of cerium trifluoroacetate in the presence of oleic acid at elevated temperatures can yield multiple products due to competing reaction pathways. The trifluoroacetate ligands decompose to release fluorine-containing species (\ce{HF}, \ce{F-}). While the reducing environment created by the alkene groups in ODE and OA is likely to preserve the Ce$^{3+}$ oxidation state, cerium oxides such as \ce{Ce2O3} could also form alongside \ce{CeF3} under these conditions. Mixed oxyfluoride phases such as \ce{CeOF} or \ce{Ce2O2F2} are also known to be stable in the bulk form owing to similar ionic radii of O$^{2-}$ and \ce{F-} anions~\cite{popov1954, udayakantha2019}. The competition between complete oxidation (leading to pure \ce{Ce2O3}) and fluoride incorporation (yielding oxyfluorides or \ce{CeF3}-like phases) depends on reaction kinetics, temperature, and the local concentration of fluoride species during nanoparticle formation. Other phases that are not stable in the bulk could also form due to the presence of large amounts of surfactants. Although Zhang \textit{et al.}  \cite{zhang_SingleCrystalline_2005} reports the formation of pure \ce{CeF3}, we show in the following that further investigation reveals the presence of oxygen in the NPL.

\begin{figure}[b!]
    \centering
    \includegraphics[width=0.95\linewidth]{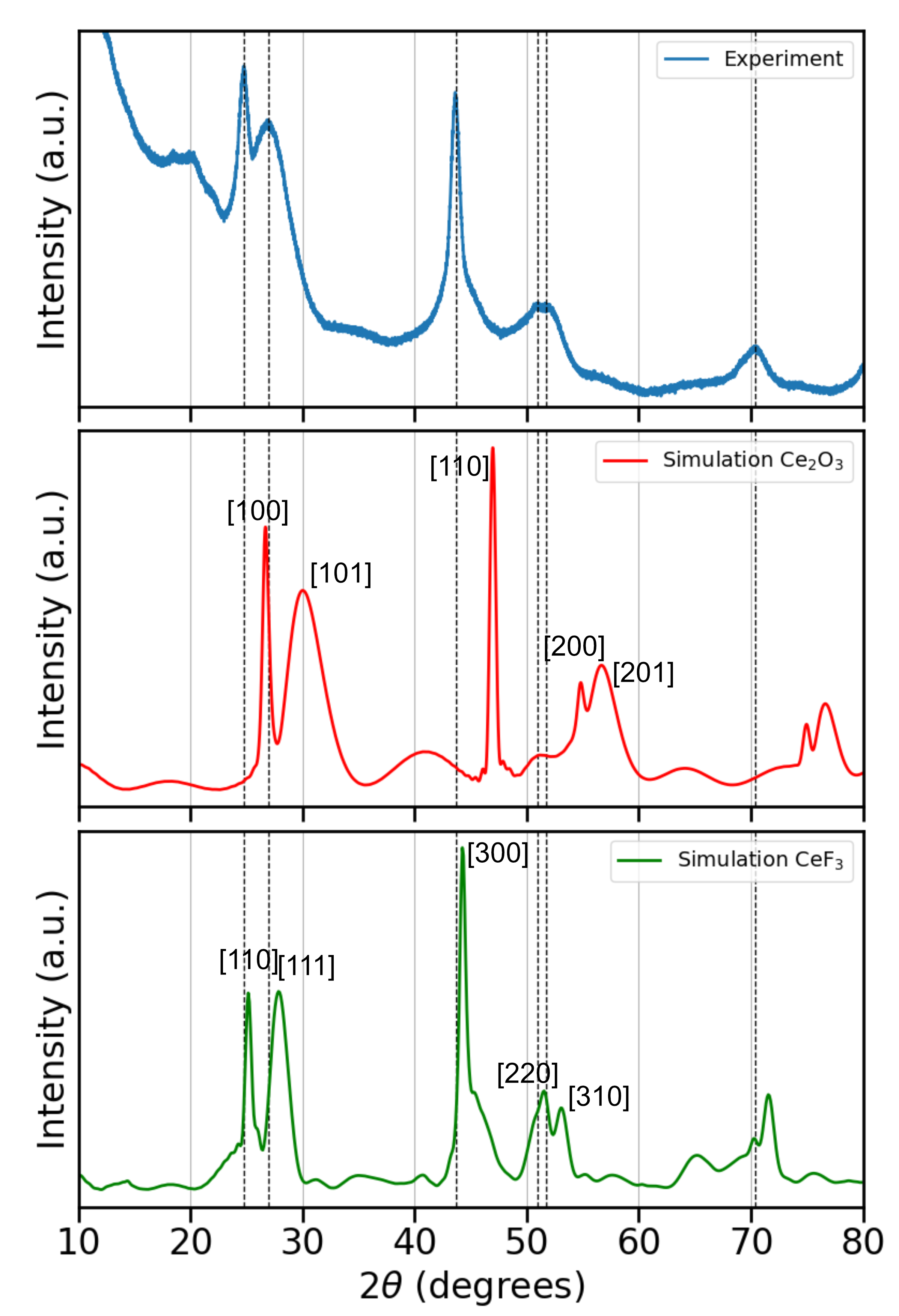} 
    \caption{Experimental powder X-ray diffraction pattern of the NPLs (blue), compared with simulated patterns for \ce{Ce2O3} (red) and \ce{CeF3} (green) NPLs.} 
    \label{fig:diffraction}
\end{figure} 

We first characterized the NPL crystal structure using X-ray powder diffraction and compared the experimental pattern with simulated diffractograms from atomic models. Triangular NPL models were constructed assuming either a \ce{Ce2O3} structure (space group P$\overline{3}$m1) or a \ce{CeF3} structure (space group (P$\overline{3}$c1)), and their powder diffraction patterns were simulated using the Debye formula (Fig. \ref{fig:diffraction}). In both cases, we assumed that the [001] axis is perpendicular to the NPL thickness, consistent with the TEM results. By simulating powder X-ray diffraction patterns directly from the atomic coordinates of a finite-size nanoparticle, the anisotropic broadening of diffraction peaks can be predicted without relying on empirical line-shape models. This approach provides a direct link between particle morphology and the experimentally observed orientation-dependent peak widths.  The simulated \ce{CeF3} pattern displays a doublet of peaks at 25.16$^{\circ}$ and 27.87$^{\circ}$. The first peak at 25.16$^{\circ}$ is sharp and corresponds to the [110] reflection, which is in plane and is less broadened by the limited extension of the NPL in the (001) direction. The second, broader peak at 27.87$^{\circ}$ can be attributed to the [111] planes. The very intense peak at 44.3$^{\circ}$ corresponds to the [300] plane, while the doublet between 50 and 55$^{\circ}$ represents the [220] and [310] planes. The same indexation for the simulated \ce{Ce2O3} yields peak attributions consistent with sharp peaks within the NPL plane and broadened reflections for non-zero $k$ Miller indices, as expected \cite{holder2019}. Interestingly, the two simulated patterns, for \ce{CeF3} and \ce{Ce2O3} display an overall shape that matches the experimental diagram: a doublet of peaks in the 20-30$^{\circ}$, a sharp intense peak between 40 and 50$^{\circ}$, and another doublet at higher angles. The first peak at around 20° in the experimental PXRD diagram (which corresponds to a q = 1.4 Å$^{-1}$) can be assigned to a harmonic of the stacking peak, which is likely to be present in a dry powder. A closer comparison between the experimental NPL diagram and the simulated ones reveals that the experimental diffraction pattern is better aligned with \ce{CeF3}, as its peak positions match those of \ce{CeF3} more closely. All the peaks in the experimental diagram can be assigned to the simulated ones, with a small shift towards higher angles, as shown in Fig. \ref{fig:diffraction}, where the dotted lines correspond to the experimental peak positions. These shifts correspond to expansive strains between 1 and 3 \%. In contrast, the peak positions predicted for \ce{Ce2O3} with bulk lattice parameters are heavily shifted to high angles compared to the experimental ones. These differences would imply strains between 7 and 11 \%. Though ligand induced stress are known to induce significant strain on NPL crystal lattices, reported strains on CdSe NPL with similar thicknesses and surface chemistry typically fall in the 2 to 5 \% range rather than higher \cite{antanovich_strain-induced_2017, Chen2015}. For spherical metallic nanocrystals, strains in the same range have been measured by electron tomography science \cite{kim2020a} with a maximal strain of 5\% for the first monolayers. Hence, assuming a \ce{Ce2O3} structure would require strains well above the range reported for comparable systems, supporting \ce{CeF3} rather than \ce{Ce2O3} as the more likely composition.

\begin{figure}[b!]
    \centering
    \includegraphics[width=0.95\linewidth]{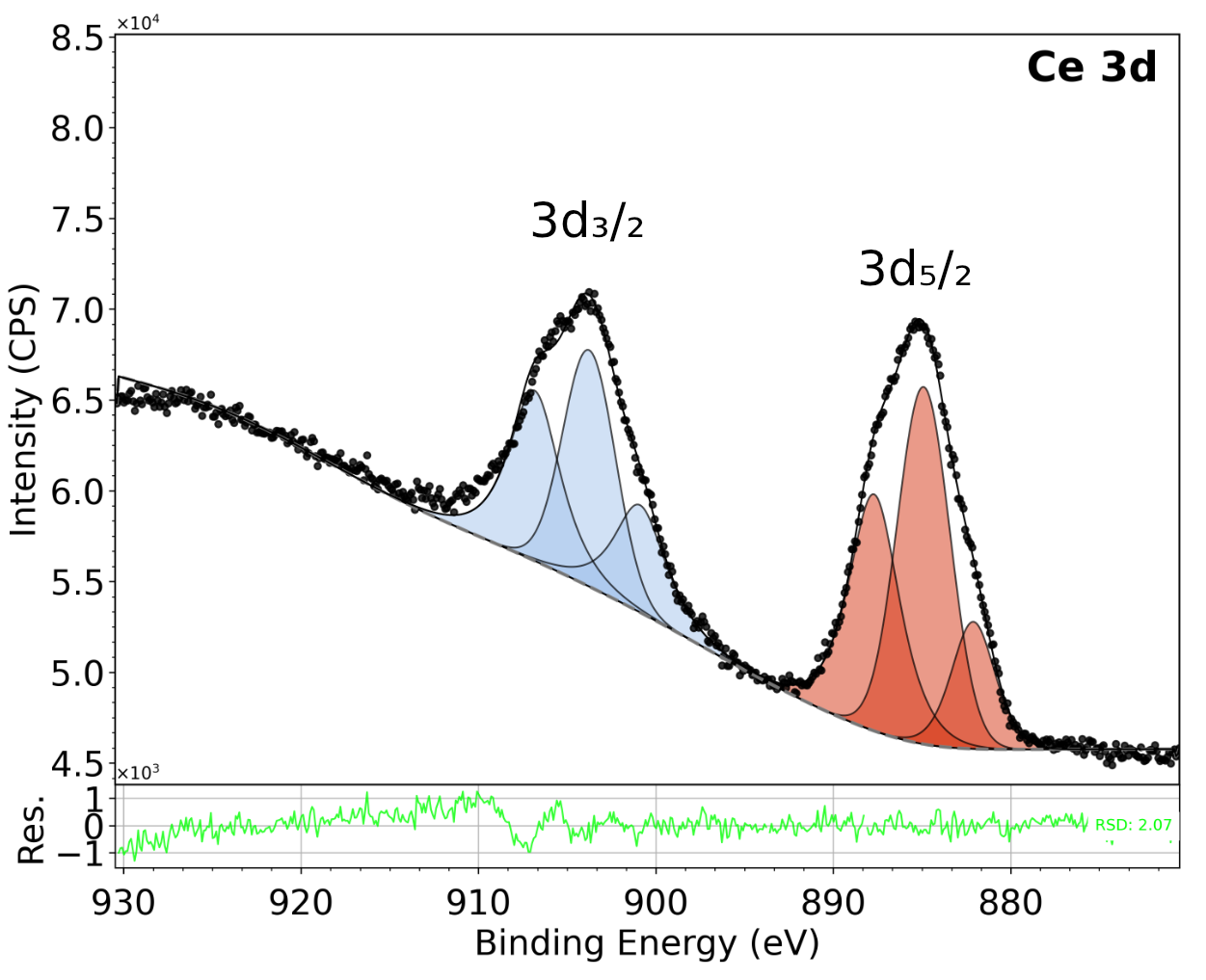} 
    \caption{3d core level XPS spectrum of the NPL. The two peaks are fitted with three peaks to obtain the experimental shape. Note the absence of peaks for energies higher than 910 eV.} 
    \label{fig:XPS}
\end{figure} 

X-ray Photoelectron Spectroscopy (XPS) brings further insight into the NPL composition. The XPS survey spectrum indicates that only C, O, Ce and F elements are present in the sample (Fig. \ref{figSI:XPS_survey}). The high resolution spectra yield the following atomic percentages (averaged for two different regions): C (86.3), O (10.0), Ce (1.6) and F(2.1). We note that the experimental C/O ratio (8.6) is lower than the theoretical one for oleic acid (9). This suggests the presence of oxygen within the crystalline cores or at the surface of the colloidal nanocrystals. This is confirmed by the measured F/Ce ratio (1.31), which is also lower than the theoretical ratio for a pure \ce{CeF3} compound (3.0). The technique sensitivity is comprised between 0.1 and 0.5 \% at and variations of 0.3 to 0.7 \% are observed depending on the region of the sample probed. The oxyfluorine nature of the crystalline core is also attested by the Ce 3d high-resolution XPS spectrum (Fig. \ref{fig:XPS}), which displays a complex shape. Previous studies \cite{park1993, creaser1994} have shown that 3d core-level XPS spectra of \ce{CeF3} and \ce{Ce2O3} display two double peaks in the 920-880 eV range attributed to the 3d$_{3/2}$ and 3d$_{5/2}$. For each double peak, the lower binding-energy peaks are higher in intensity in the case of \ce{CeF3} \cite{park1993} but lower for \ce{Ce2O3} \cite{creaser1994}. The positions of these peaks also subtly vary depending on the cerium coordination within the 880-890 eV window for 3d$_{3/2}$ and 900-910 eV for 3d$_{5/2}$. We indeed observe two peaks in the same energy region, but the peak lineshapes differ from the ones observed previously for \ce{CeF3} and \ce{Ce2O3}. If only two peaks are considered for each energy level, the Ce 3d peaks cannot be fitted satisfactorily; an additional peak must be included to obtain the correct lineshape. We also note the absence of a peak at energies between 910 and 920 eV, whereas cerium at the +4 oxidation state typically displays a strong peak in this energy range \cite{creaser1994}. Hence, our XPS results establish that the NPLs contain cerium-bound oxygen beyond the ligand contribution, consistent with an oxyfluoride composition

Thermogravimetric analysis (TGA) of a NPL sample shows a significant weight loss in the 25-500 $^{\circ}$C range (Fig.\ref{figSI:TGA}). In air, 71.54 \% of the initial mass is lost while a smaller fraction (67.78 \%) is lost in an inert \ce{N2} atmosphere. Interestingly, the chemical nature of the residue depends on the atmosphere under which the decomposition occurs. The PXRD of the residue (Fig \ref{figSI:TGA}B) in air can be unambiguously assigned to \ce{CeO2}, showing that, during heating, the initial NPLs undergo complete oxidation with total fluorine elimination. In contrast, when heating occurs under an inert atmosphere, the residue includes oxide and fluoride compounds whose PXRD diagram can not be indexed to either \ce{CeO2} or \ce{CeF3}. The difference in weight loss between the two atmospheres enables an estimation of the F and O content of the NPLs. By making reasonable hypothesis on the degradation process (see SI for details) we find that the NPL are fluorine rich with a Ce:F atomic ratio of 1:2.7. Since the XPS spectrum of \ce{Ce} discards the presence of \ce{Ce^{IV}}, the NPL have a mean chemical formula of \ce{CeO_{0.15}F_{2.7}}. We stress that the oxygen content measured here may be underestimated and should be treated as a lower bound, as oxygen in the NPL core could be released upon reaction with organic ligands at temperatures below 500 $^{\circ}$C. \\

We also investigated the NPL crystalline lattice structure using high-resolution transmission electron microscopy (HRTEM). Figure \ref{fig:TEM_diffraction}b shows an image of a triangular flat NPL with the corresponding Fast Fourier Transform (FFT). Two series of reflections arranged in a hexagonal pattern indicate a hexagonal crystallographic structure. The first-order peaks (first regular hexagon) are at 0.335 nm on average, and the second-order peaks are at 0.189 nm on average. We simulated selected-area electron diffraction patterns of trigonal \ce{CeF3} and \ce{Ce2O3} NPL along the (001) direction (Fig. \ref{fig:SAED}). The first one has a P$\bar{3}$c1 structure (a = b = 7.0850 Å and c = 7.2397 Å, reference mp-510560 in Materials Project \cite{jain2013}) and the second a P$\bar{3}$m1 with a = b = 3.87 Å and c = 5.90 Å, reference mp-2721). In \ce{CeF3}, the first-order reflections are visible at 0.61358 nm while two sets of higher-order reflections are simulated at 0.3542 and 0.20543 nm, respectively. We note that, in this case, the first low-index reflections corresponding to the [100] family have much lower intensities, which could explain their absence in the experimental FFT.  In contrast, the first two sets of reflections for \ce{Ce2O3} appear at 0.33524 and 0.19355 nm, respectively, i.e., very close to the experimental distances. Most triangles exhibit polycrystallinity. The large triangular facet corresponds to the (001) plane, and the edges to the (100), (0-10), and (-110) planes. NPL surfaces are not atomically flat: as seen on the HRTEM images of NPLs that lie perpendicular to the substrate (Fig.~\ref{fig:Fig1METSA}c and d), steps are observed on the (001) surfaces on individual NPLs. Most NPLs observed consist of four monolayers, with a small minority of three and five monolayers. We observe that the NPL structure is affected by the electron beam during imaging. Irradiation increases surface roughness, and even if atomic steps are already visible in the first images of the NPLs, they can create defects and alter the lattice parameters. Figure \ref{figSI:beameffect} shows two pictures of the same area of the TEM grid at time 0 and after 5 minutes of exposure. Comparing these two images shows that the roughness, edges, and crystalline structures of the NPLs are affected. Hence, it is difficult to disentangle the beam effect from pre-existing defects, and the roughness cannot be analyzed quantitatively. Finally, STEM-EDX analysis (Fig. \ref{figSI:EDX}) confirms that cerium is localized within the NPLs and is not detected in other regions of the TEM grid, whereas fluorine and oxygen are found across all probed areas. The ubiquitous fluorine signal is attributed to residual fluorinated species originating from the precursor, which are redistributed over the grid during drop-casting. Importantly, the F/O ratio is higher on the NPLs than on bare grid regions. Together with the co-localization of cerium and fluorine in the NPLs, this supports the incorporation of fluoride into the NPL.

Overall, the detailed investigation using several complementary techniques indicates that the synthesized NPLs have an oxyfluoride composition, with both oxygen and fluorine present. While electron diffraction patterns are more consistent with a \ce{Ce2O3} structure, this would imply very large strains ($\simeq 10 \%$), as shown by X-ray diffraction. Both XPS and TGA results are consistent with a mixed compound with a generic chemical formula CeO$_x$F$_y$. Determining $x$ and $y$ precisely would require further experimental efforts. An open and interesting question concerns the timing of oxygen incorporation into the NPL. Several plausible scenarios exist: oxidation may occur during synthesis itself, given the harsh conditions (excess oleic acid, high temperature); alternatively, pure \ce{CeF3} NPL may form initially and oxidize gradually during storage and handling under air. There is also a possibility that oxidation is accelerated upon drying on a surface, which would naturally explain the higher oxidation levels observed under STEM conditions. Clarifying these pathways would require systematic air-free handling at each stage of synthesis, purification, and characterization.

\subsection{NPL self-assembly in solution}
\begin{figure}[t!]
    \centering
    \includegraphics[width=0.95\linewidth]{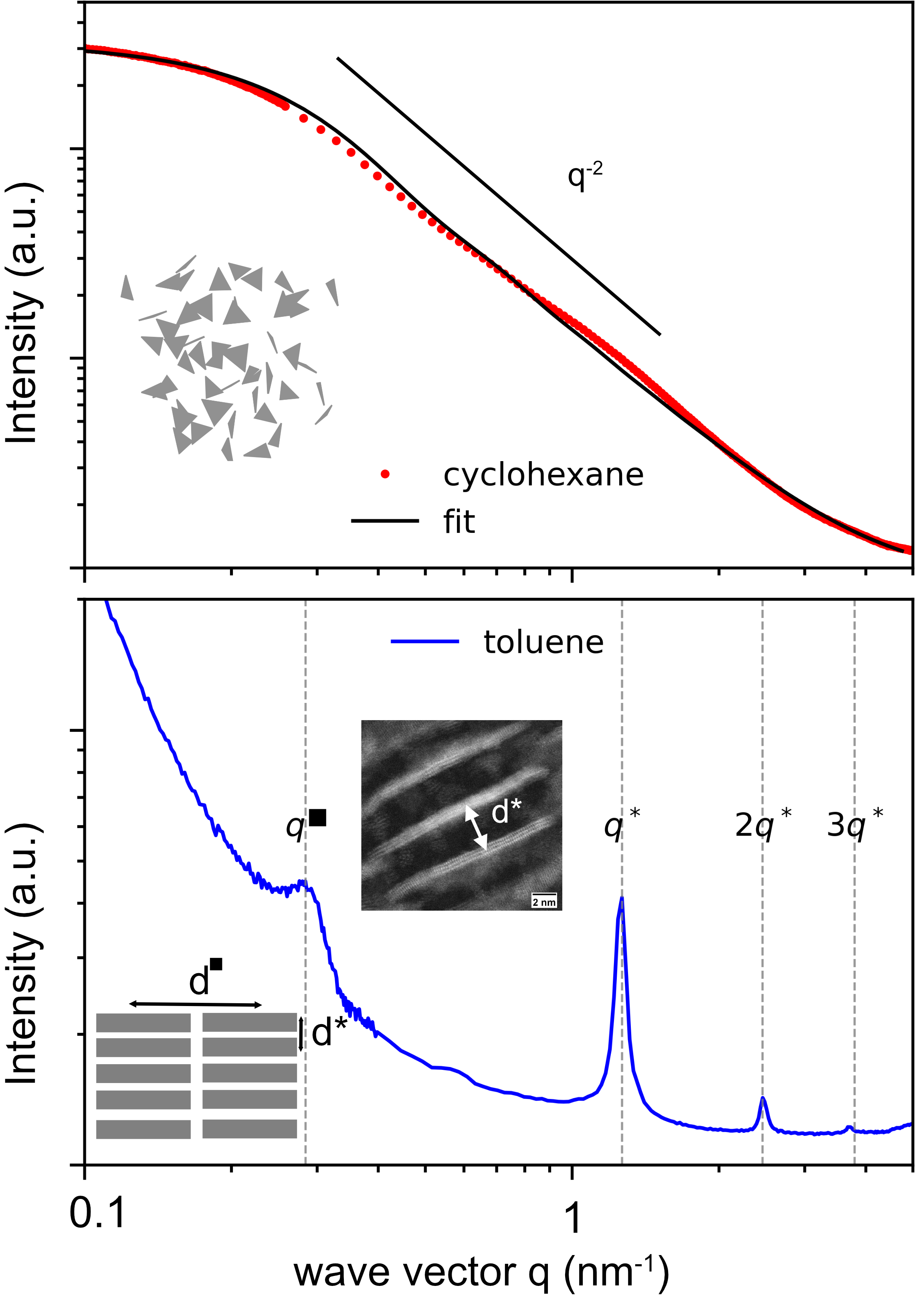} 
    \caption{Top: SAXS pattern of cerium NPLs dispersed in cyclohexane (red line), and the fitting (black line) with a model of prismatic particle with 3 edges. Insert: schematic illustration of randomly oriented NPLs in solution. Bottom: SAXS pattern of cerium NPLs dispersed in toluene. The peaks are at q$^{-1}$ positions of 0.284, 1.265, 2.460 and 3.676 nm$^{-1}$. Inserts: STEM image of triangular NPLs stacked face-to-face and schematic illustration of NPL stacking.} 
    \label{fig:stacking}
\end{figure} 

We explored the effect of the solvent on the self-assembly of these triangular NPLs in solution using Small Angle X-ray Scattering (SAXS), a well-suited technique to gain structural insight on nanoparticles and their assemblies in solution \cite{guinier_small-angle_1955, Als-Nielsen2001, li2016, jeffries2021}. To do so, we dispersed NPLs with an average edge length of 16.3 nm in two nonpolar solvents: toluene and cyclohexane. The SAXS patterns of the two dispersions differ markedly. In cyclohexane, the SAXS pattern is monotonously decreasing with a slope at intermediate $q$ close to $q^{-2}$ (Fig.~\ref{fig:stacking}). This is consistent with flat objects dispersed in solution. Fitting the experimental scattering diagram with a model of prismatic particle with 3 edges \cite{Marcone2015} yields a reasonable agreement for a mean edge value of 18.7 nm and a polydispersity of 5 \% and a thickness of 0.7 nm, very close to values measured by TEM. We note that the quality of the fit does not vary with the edge polydispersity. In toluene, the SAXS pattern shows four peaks (Fig.~\ref{fig:stacking}). The more intense peak appears at q* = 1.265 nm$^{-1}$. Two other peaks appear at 2q* = 2.460 nm$^{-1}$ and at 3q* = 3.676 nm$^{-1}$, which indicates that the NPLs are stacked one on top of another face-to-face in a lamellar fashion, where $d=2\pi/q^*$ = 5 nm corresponds to the repeating periodicity between NPLs. This matches closely with the sum of twice the length of the ligands (2x2 nm for OA) plus the thickness of the NPL (0.7 nm), meaning that the ligands are not interdigitated. This value is slightly larger than the value measured by STEM (4.1 nm). The difference can be attributed to the different sample environments: in SAXS, the NPLs are dispersed in a solvent, which can occupy the space between the NPLs and penetrate the ligand shell, while the NPLs analyzed by TEM or STEM have been previously dried under a vacuum on a solid support. At lower wave-vectors, a less intense peak at q$^{\blacksquare}$ = 0.284 nm$^{-1}$, corresponding to a distance of 22.1 nm, appears in the toluene SAXS pattern. This peak can be attributed to lateral stacking of NPL columns within a 2D liquid-like structure (Fig.~\ref{fig:stacking}). Considering the NPL edge dimension, the lateral NPL edges are thus separated by 6 nm on average, implying the presence of solvent and no ligand interdigitation in the gap within the layers. 

\begin{figure}[! h]
    \centering
    \includegraphics[width=0.95\linewidth]{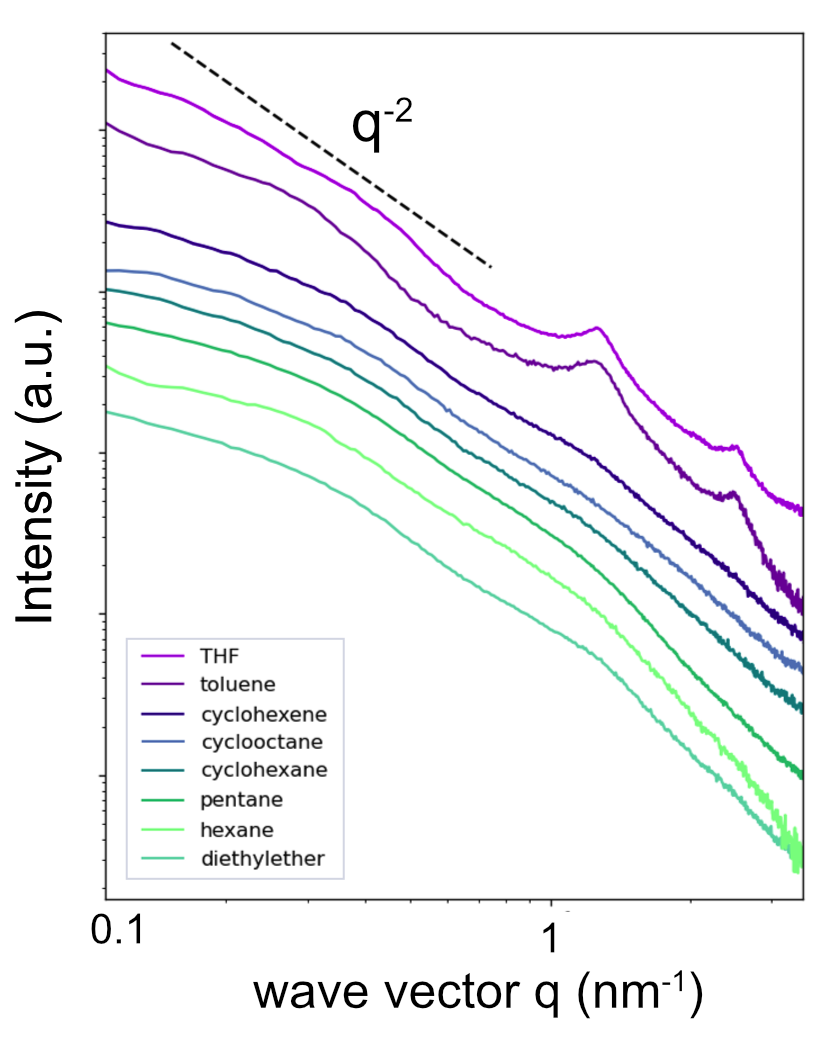}
    \caption{SAXS analysis of cerium NPLs dispersed in different organic solvents. Note the $q^{−2}$ slope showing the 2D character of the scattering objects.} 
    \label{fig:SAXS-solvent_slope}
\end{figure}

Beyond toluene and cyclohexane, we expanded the range of organic solvents to cyclohexene, cyclooctane, diethyl ether, \textit{n}-hexane, \textit{n}-pentane, and THF. The stacking peaks are only visible for toluene and THF. In all the other solvents, the SAXS patterns decrease monotonously with a q slope close to q$^{-2}$ as for cyclohexane (Fig.~\ref{fig:SAXS-solvent_slope}). To rationalize these solvent effect, we resort to the Flory-Huggins theory which has been used previously in a similar context \cite{raghavan2000, khan2009, ofosu2022,monego_when_2020}. In this framework, the interaction parameter $\chi$ characterizes the affinity for a given ligand-solvent pair: $$\chi=\frac{V_s}{RT}(\delta_s - \delta_l)^2+0.34$$, where $\delta$ correspond to Hildebrand solubility parameters of the solvent ($s$) and ligand ($l$) respectively and $V_s$ is the specific volume of the solvent. The first term in the right-hand side of the equation corresponds to the enthalpic contribution of the mixing to which adds an empirical constant (0.34) for the entropy of mixing. For polymers with infinite number of monomers, solutions are stable for $\chi$<0.5 and unstable for $\chi$>0.5. In this framework, solutes that have Hildebrand solubility parameters close to the one of the solvent minimize the enthalpic term, meaning that "like dissolve like". Table \ref{tab:chi_solvents} shows the $\chi$ values for the different solvents for $\delta_l=16$ MPa$^{1/2}$ corresponding to the long chain alkane of the ligands \cite{brandrup1999}.

\begin{table}[htbp]
\centering
\begin{tabular}{lccc}
\hline\hline
Solvent & $V_\mathrm{s}$ & $\delta_\mathrm{T}$ & $\chi$ \\
 & (cm$^3$\,mol$^{-1}$) & (MPa$^{1/2}$) & \\
\hline
Toluene        & 106.8 & 18.2 & 0.55 \\
THF            & 81.7  & 19.5 & 0.74 \\
Cyclohexene    & 101.9 & 17.4 & 0.42 \\
Cyclohexane    & 108.7 & 16.8 & 0.37 \\
Pentane        & 116.2 & 14.5 & 0.45 \\
Hexane         & 131.6 & 14.9 & 0.40 \\
Diethyl ether  & 104.8 & 15.5 & 0.35 \\
\hline\hline
\end{tabular}
\caption{Hildebrand total solubility parameters $\delta_\mathrm{T}$, molar volumes $V_\mathrm{s}$, 
and Flory-Huggins interaction parameters $\chi$ of the solvents with respect to the alkyl ligand shell of the nanoplatelets. The values are taken from reference \cite{brandrup1999}.}
\label{tab:chi_solvents}
\end{table}

The computed $\chi$ values correlate well with the observed behavior of the NPL in these solvents: dispersion is obtained for all solvents with $\chi$<0.5, whereas stacking occurs for the only two solvents with $\chi$ > 0.5 (toluene and THF). This simple solvation picture, in which the NPL is treated as long alkyl chains, therefore accounts well for our experimental observation.

This SAXS analysis in solution demonstrates strong solvent effects consistent with the conventional "like dissolves like" principle, in which apolar solvents impart greater colloidal stability than more polar solvents, such as THF or toluene \cite{monego_when_2020}.

For selected solvents, we investigated the NPL self-assembly at the liquid-air interface by following a method first described by Dong \textit{et al.} \cite{dong2010}. The dispersions are spread onto a liquid surface (DEG), and the solvent evaporates slowly. Depending on the solvent, the resulting NPL arrays are arranged in several ways: lying flat on the face (edge-to-edge) to form hexagonal close-packed (\textit{hcp}) superlattices, face-to-face on the edge to assemble into stacks, or without a long-range order arrangement. 

\begin{figure*}
    \centering
    \includegraphics[width=0.6\linewidth]{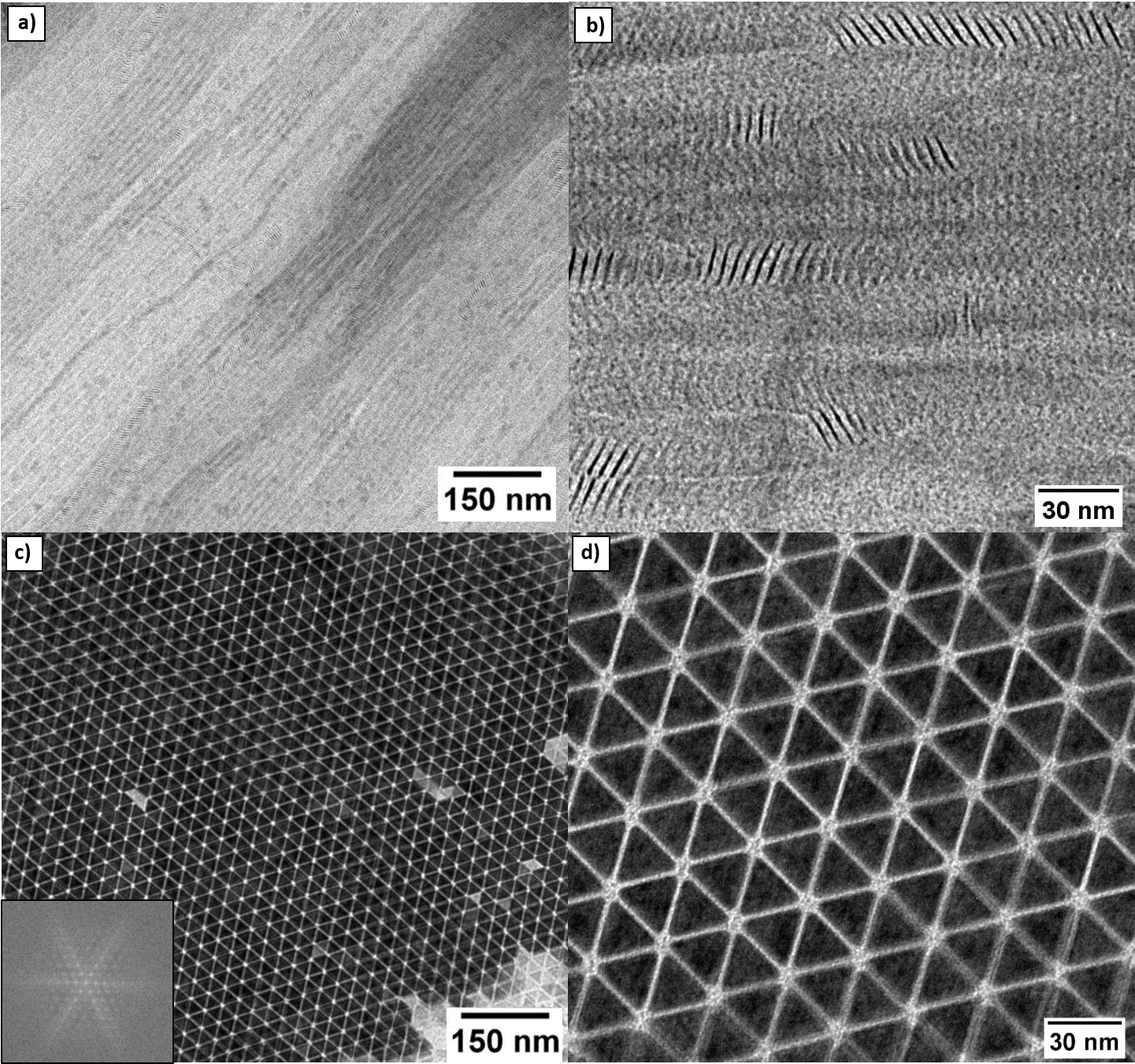}
    \caption{TEM images illustrating the NPL self-assembly after the evaporation in toluene (a,b) and cyclohexane (c,d) on DEG.} 
    \label{fig:assembly_tol_cyclo}
\end{figure*}

Films obtained from toluene dispersions display long-range arrays of NPLs stacked in an edge-up configuration. The stacks span from hundreds of nanometers to tens of micrometers with NPL columns arranged parallel (Fig.~\ref{fig:assembly_tol_cyclo}a and b), consistent with the SAXS results and confirming the formation of long, columnar, adjacent stacks of NPLs that were already present in solution before evaporation. Remarkably, the arrays are parallel, which explains the peak at low q, corresponding to the intercolumn spacing. TEM images show a spacing between adjacent stacks of around 2 nm or less, smaller than the distance measured by SAXS (5 nm), due to the absence of solvent. In contrast, when the NPLs are dispersed in cyclohexane, they assemble at the liquid-air interface into long-range periodic superlattices covering areas of tens of micrometers. However, in this case, the NPLs are arranged in a face-down, edge-to-edge configuration (Fig.~\ref{fig:assembly_tol_cyclo}c,d and Fig.~\ref{figSI:STEM-SA-cyclohexane}). The resulting films consist of periodic superlattices with hexagonal close-packed symmetry, composed of equilateral triangles with an average interparticle distance of about 2 nm between neighboring NPLs. These superlattices span several micrometers. By varying the NPL concentration, hexagonal close-packed multilayers form, with NPLs perfectly stacked face-to-face and edge-to-edge. (Fig.~\ref{fig:assembly_tol_cyclo}c.) The layer thickness can be inferred from the image contrast: the darkest areas in the TEM image correspond to regions where fewer electrons are transmitted through the sample because of greater thickness, whereas lighter regions correspond to thinner zones. In the bottom right of Fig. \ref{fig:assembly_tol_cyclo}c we can notice an area composed of a single monolayer, while the darkest areas comprise three layers.

A different trend is observed from the self-assembly of NPLs dispersed in all the other solvents (hexane, cyclohexene, cyclooctane, diethyl ether, and pentane). Instead of close-packed hexagonal structures or long-range stacks, TEM images show films composed of regions with varying structures, mainly in a face-down configuration with only a few face-to-face arrays observed in cyclohexene and cyclooctane samples (Fig.~\ref{figSI:TEM-SA-cyclohexene,cyclooctane}). In THF, the NPLs are preferentially stacked face-to-face into small arrays but without long-ranged order (Fig.~\ref{figSI:TEM-SA-THF}) as for toluene. A few areas of the TEM grid show edge-to-edge organizations (Fig.~\ref{figSI:TEM-SA-THF}).
For hexane, Fig. \ref{fig:TEMCe2O3hexane}a shows NPLs in a face-down assembly with a loosely organized structure. Although there is no orientation order as for cyclohexane, the FFT shows rings, indicating some short-range positional order. Such a "glassy" structure is intermediate between a liquid and a crystal, characterized by long-range positional and orientational order, and it suggests a kinetically trapped state in which the particles did not have sufficient time to orient before the solvent was completely evaporated. In other parts of the TEM grid, the NPLs were stacked in an edge-up configuration without forming long stacks (Fig.~ \ref{fig:TEMCe2O3hexane}b.). Our experiments thus show that the post-evaporation structure arises from both pre-existing assemblies in solution and kinetic effects during evaporation. As Momper \textit{et al.} have shown, rapid evaporation yields face-down kinetically trapped assemblies \cite{momper_kinetic_2020-1}. In contrast, slow evaporation enables the NPL to rearrange and ultimately yield edge-up assemblies with stacked NPLs. In our case, we indeed observe that toluene, the solvent that evaporates the slowest (see Table \ref{tabSI:solvents} for the vapor pressures of the different solvents), yields long-range assemblies with face-to-face stacks. While THF also induces stacking in solution, the solvent's rapid evaporation yields assemblies with short-range order. Similarly, NPL that are well-dispersed in solution yield long-range edge-to-edge structures only when the solvent evaporates slowly, like in cyclohexane. Faster evaporation (hexane, pentane) only affords glassy structures. 

\section{Conclusions} 
In conclusion, we report an optimized protocol to yield triangular cerium-based NPLs with low polydispersity in size and shape. These NPLs were extensively characterized using HR-STEM, electron diffraction, XRD, XPS, and TGA, revealing partial oxidation of the expected \ce{CeF3} structure into an oxyfluoride composition (\ce{CeOxFy}). We demonstrate that the NPLs exhibit strong solvent-dependent assembly behavior, both in solution and upon evaporation at the liquid-air interface. Depending on the solvent, NPLs assemble into well-dispersed individual particles, long-range face-to-face stacks, or extended hexagonal close-packed superlattices. Our results establish clear correlations between solvent properties, evaporation kinetics, and the resulting NPL organization, providing guidelines for controlling the collective arrangement of 2D nanostructures. These insights can be leveraged to rationally design NPL-based materials.

\section{Experimental Methods}

\subsection{Chemicals} 
~~~ \ce{Ce(CF3COO)3} hydrate was purchased from Jiayuan.
Oleic acid (90\%), cyclohexane, and cyclooctane were purchased from Sigma-Aldrich. 
Octadecene (90\%), n-hexane (95\%), n-pentane, and acetone were purchased from Fisher-Acros Organics.
Diethylene glycol (DEG, 99\%), diethyl ether, and THF were purchased from Alfa Aesar. 
Ethanol and toluene (99.5\%) were purchased from VWR.

\subsection{Synthesis of triangular cerium-based NPLs}
\ce{Ce(CF3COO)3} hydrate (1 mmol, 479 mg),  oleic acid (31 mmol, 10 mL), and octadecene (31 mmol, 10 mL) were mixed in a 50 mL three-neck flask. The mixture was placed under vacuum and heated to 100 °C for 30 minutes to remove water and oxygen, under vigorous magnetic stirring in a temperature-controlled electromagnetic mantle. White foam and bubbles formed during the degassing period. The white precursor was dissolved, yielding a light orange solution. The flask was then placed under argon flux, heated to 260 °C at 18 °C/min using a heating mantle (Pilz, WHG-Classic model WHG2, 1000 ml, 450 W) controlled by a temperature regulator (Chauvin Arnoux, Statop 48 model) and held at this temperature for 1 hour. A light-yellow colloidal solution was obtained and was cooled to room temperature. The NPLs were washed by adding a mixture of hexane (10 mL)/acetone (40 mL) into the colloidal solution and precipitated by centrifugation (5000 rpm, 5 min). The precipitate was then washed with ethanol, sonicated, and centrifuged. These purification steps were repeated twice. The typical reaction yield is 70 \%, evaluated by weighing the powder after purification and accounting for the ligand contribution determined by TGA. The as-prepared NCs were then redispersed in hexane. 

\subsection{Self-assembly of triangular NPLs in solution}
To prepare the glass capillaries for SAXS analysis of the NPL dispersions in nonpolar solvents, NPLs were purified as follows.  NPLs in hexane (5 mg/mL) were precipitated by adding 20 mL of acetone, followed by centrifugation (5000 rpm, 5 min). The supernatant was discarded, and the white precipitate was redispersed in 5 mL of hexane and 5 mL of ethanol and sonicated for 10 minutes. A yellow precipitate was recovered after centrifugation (3000 rpm, 3 min) and redispersed in 8 mL of hexane. Four 15 mL centrifuge tubes were filled with 2 mL each of this dispersion, to which 8 mL of acetone was added, and then centrifuged at 5000 rpm for 5 min. The supernatants were removed, and the white precipitates were redispersed in 2 mL of a nonpolar solvent (cyclohexane, cyclohexene, cyclooctane, diethyl ether, n-hexane, n-pentane, THF or toluene). Special glass capillaries (1 mm diameter, CTS Capillary Tube Supplies Ltd) were filled with these dispersions (20 mg/mL) and analyzed by SAXS. 

\subsection{2D self-assembly at the liquid-air interface}
2D self-assembly experiments were carried out using a drying-mediated method \cite{dong2010}. A Teflon well with a cylindrical hole (1 cm in diameter and height) filled with 0.3 mL of DEG was used as the support. 
20 $\mu$L of NPL dispersions in nonpolar solvents (5 mg/mL) were spread on the liquid sub-phase in the Teflon well. The well was covered with a glass slide to slow evaporation. After complete solvent evaporation, the formed films were carefully transferred onto a TEM grid by gently placing the grid underneath the floating film using tweezers and slowly lifting it upwards to collect the assembly. The self-assembly experiments were performed at room temperature under ambient atmosphere in a fume hood. The substrates were dried under vacuum for 24 hours to remove residual DEG. The films were characterized by TEM.

\subsection{Electron microscopy}
TEM images were obtained by using a JEOL JEM2100 equipped with a LaB6 thermionic electron gun and a High Tilt objective pole piece. The microscope was operated at 200 kV with a point-to-point resolution of 0.25 nm.
HR-STEM images were acquired using a JEM-ARM200F Cold FEG NeoARM operating at 200kV at the Centre Électronique de Microscopie Stéphanois in Saint-Étienne. HRTEM images were acquired using an aberration-corrected JEOL ARM200F microscope at the MPQ lab in Paris \cite{alloyeau2012}.
ABFS filter was applied on HRTEM images to improve the contrast of thin NPL using Digital Micrograph software (Gatan inc.). The electron diffraction patterns were simulated using Crystalmaker and Single Crystal (CrystalMaker Software Ltd, version 4). 

\subsection{Small Angle X-ray Scattering}
SAXS measurements of NPLs dispersed in cyclohexane and toluene were performed on the SWING beamline at the SOLEIL synchrotron (Saint-Aubin, France) using an X-ray energy of 16 keV with a sample-to-detector distance of 1 m. SAXS measurements of NPLs dispersed in cyclohexane, cyclohexene, cyclooctane, diethyl ether, \textit{n}-hexane, \textit{n}-pentane, THF and toluene (20 mg/mL) were performed using a Xenocs Xeuss 3.0 SAXS with a Cu K$_{\alpha}$ and a sample-to-detector distance of 350 mm. The 2D patterns were radially averaged using using instrument specific procedures. The modeling and fitting of the SAXS pattern was performed using SASview \cite{krzywon2025}.

\subsection{Powder X-ray diffraction}
Powder X-Ray Diffraction (PXRD) analysis was performed with a Panalytical Empyrean X-ray diffractometer equipped with a 1.8 kW Cu K$_{\alpha}$ ceramic tube and a PIXcel3D 2x2 area detector, operating at 45 kV and 40 mA.

\subsection{Simulation of X-ray Powder Diffraction}
Atomic models of cerium oxide NPL were generated using the Crystal Maker software. The structure was then exported as an xyz coordinate text file, which serves as the input for the calculation of the X-ray diffractogram using the Debyer program (https://github.com/wojdyr/debyer) which calculates the scattering pattern from atomic coordinates using Debye’s scattering equation.

\subsection{Thermogravimetric analyses}
Thermogravimetric analyses (TGA) were carried out using a Setaram Labsys Evo TG DTADSC+ 1600 °C under either air or N$_2$ atmosphere. Samples of 30 mg were weighed in 100 $\mu$L
aluminum oxide crucibles. Before measurements, the crucibles were
pyrolyzed at 1200 °C and stored in an oven at 175 °C to ensure no
water adsorption. Then, samples were heated from RT up to 1000 °C with a heating rate of 5 K·min$^{−1}$.

\subsection{X-ray Photo-electron Spectroscopy} 
X-ray photoelectron spectroscopy (XPS) spectra were recorded on a THERMO K-alpha+ spectrometer with the Al K$\alpha$ line used as the excitation source at the Laboratoire Science et Surface of Ecully (Serma Technologies).

\begin{acknowledgements}
This work is funded by the ANR grant ANR18-CE09-0025 (Project SoftQC). This work was supported by the LABEX iMUST of the University of Lyon (ANR-10-LABX-0064), created within the "Plan France 2030" set up by the French government and managed by the French National Research Agency (ANR).
This article is also part of a project funded by the European Research Council (ERC) under the European Union's Horizon 2020 research and innovation program (Grant agreement No. 865995, ERC CoG SENECA). We thank Nicholas Blanchard and ILMTech for access to JEOL 2100 TEM and the Consortium Lyon St-Etienne de Microscopie (CLYM) for the microscope platform TEM characterisation.
The authors acknowledge the CNRS network "Microscopie Electronique et Sonde Atomique" (METSA, FR CNRS 3507) for financial support, which provided access to the aberration-corrected TEM at the MPQ lab. We also acknowledge SOLEIL for providing synchrotron radiation facilities, and we thank Thomas Bizien for assistance with the SWING beamline.

\end{acknowledgements}

\bibliography{biblio}

\onecolumngrid
\newpage

\renewcommand{\thefigure}{S\arabic{figure}}
\renewcommand{\thetable}{S\arabic{table}}
\setcounter{figure}{0}
\setcounter{subsection}{0}

\section*{Supplementary Material}

\subsection{Supplementary Figures}

\begin{figure}[htbp] 
    \centering
    \includegraphics[width=0.85\linewidth]{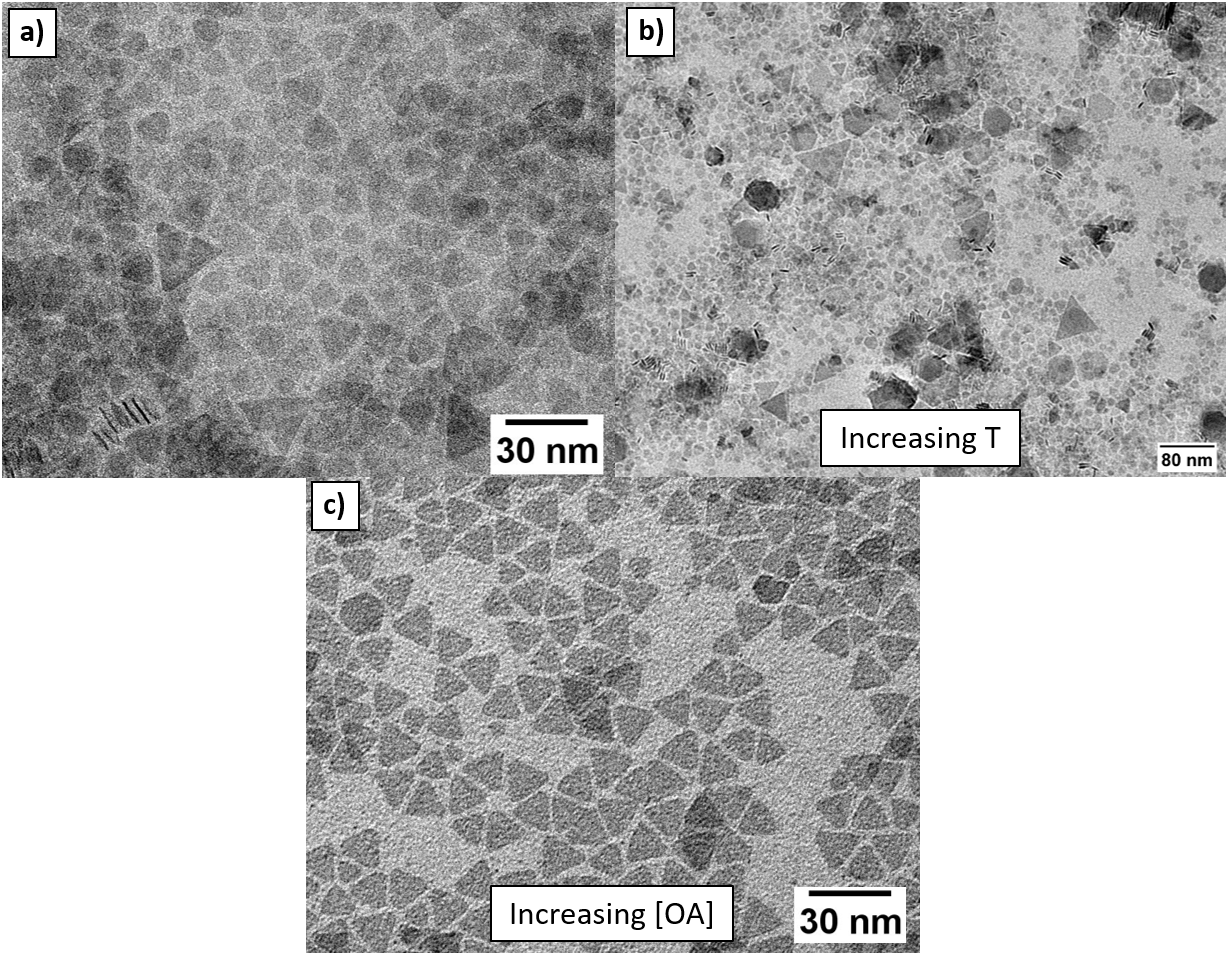}
    \caption{TEM images of triangular cerium-based NPLs obtained from different experimental parameters. a) Following the protocol reported by Zhang \textit{et al.} \cite{zhang_SingleCrystalline_2005}. b) Result of increased reaction temperature. c) Result of increased OA concentration.} 
    \label{figSI:TEM_1}
\end{figure}

\begin{figure}[htbp] 
   \centering
    \includegraphics[width=0.65\linewidth]{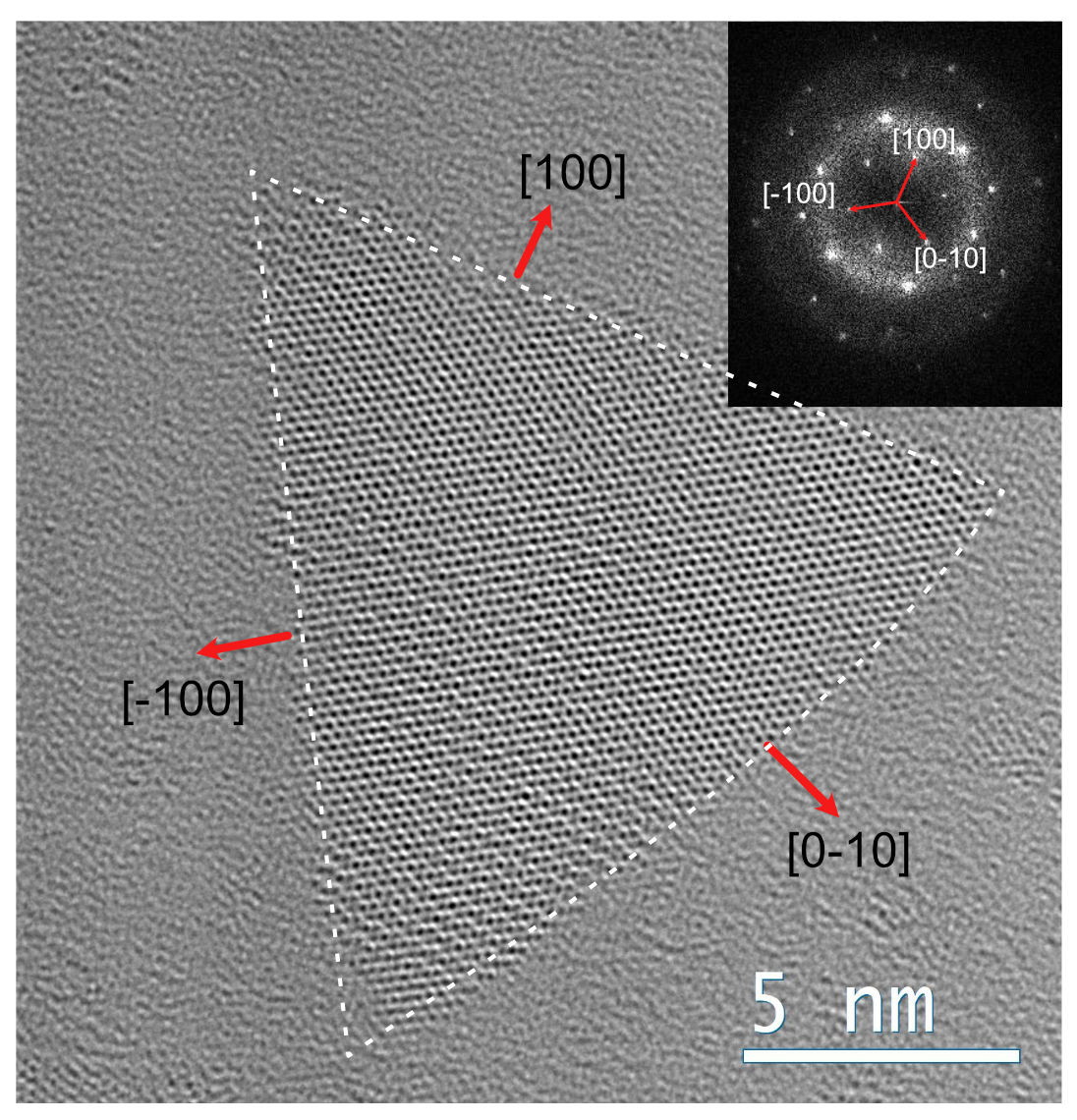}
    \caption{Average Background Subtraction Filtered (ABSF) HRTEM image and its FFT (inset) of a purified triangular NPL which contours are highlighted.} 
    \label{fig:TEM_diffraction}
\end{figure} 

\begin{figure}[htbp] 
   \centering
    \includegraphics[width=0.95\linewidth]{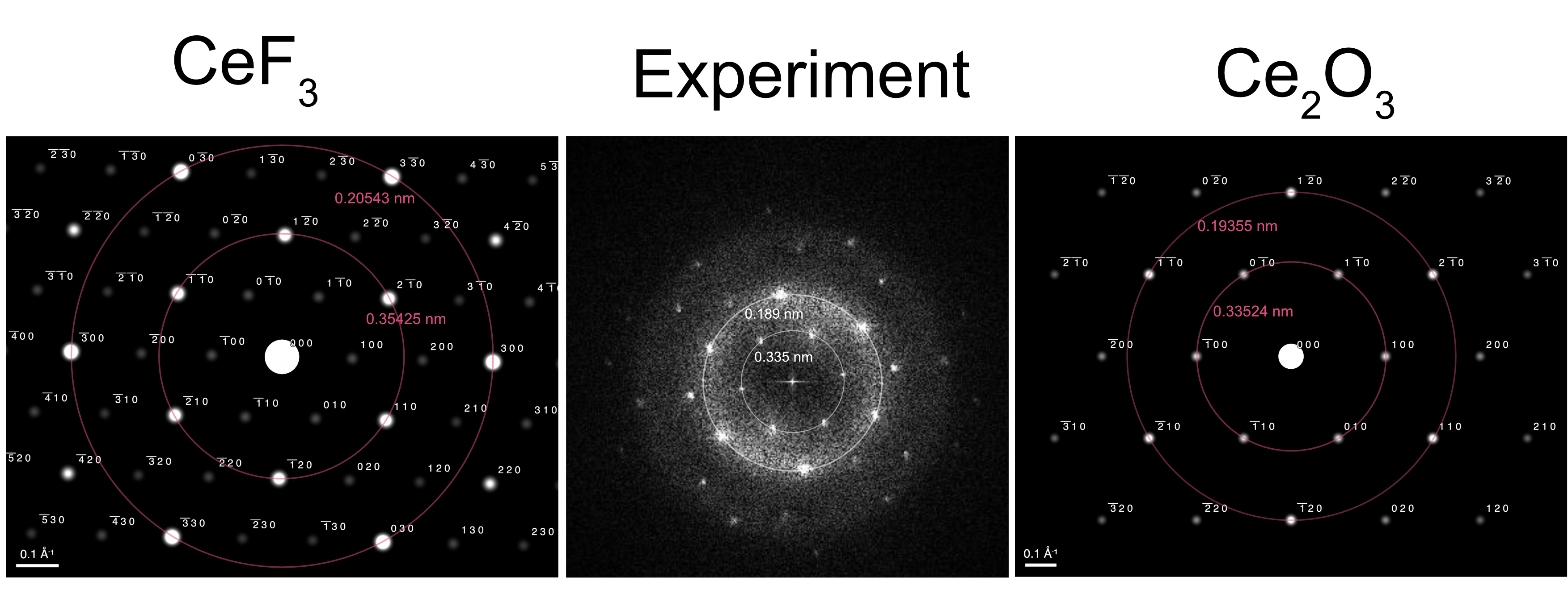}
    \caption{FFT of a high resolution image NPL (center) compared to simulations of \ce{CeF3} and \ce{Ce2O3} along the (001) zone axis.} 
    \label{fig:SAED}
\end{figure}

\begin{figure}[htbp] 
   \centering
    \includegraphics[width=0.9\linewidth]{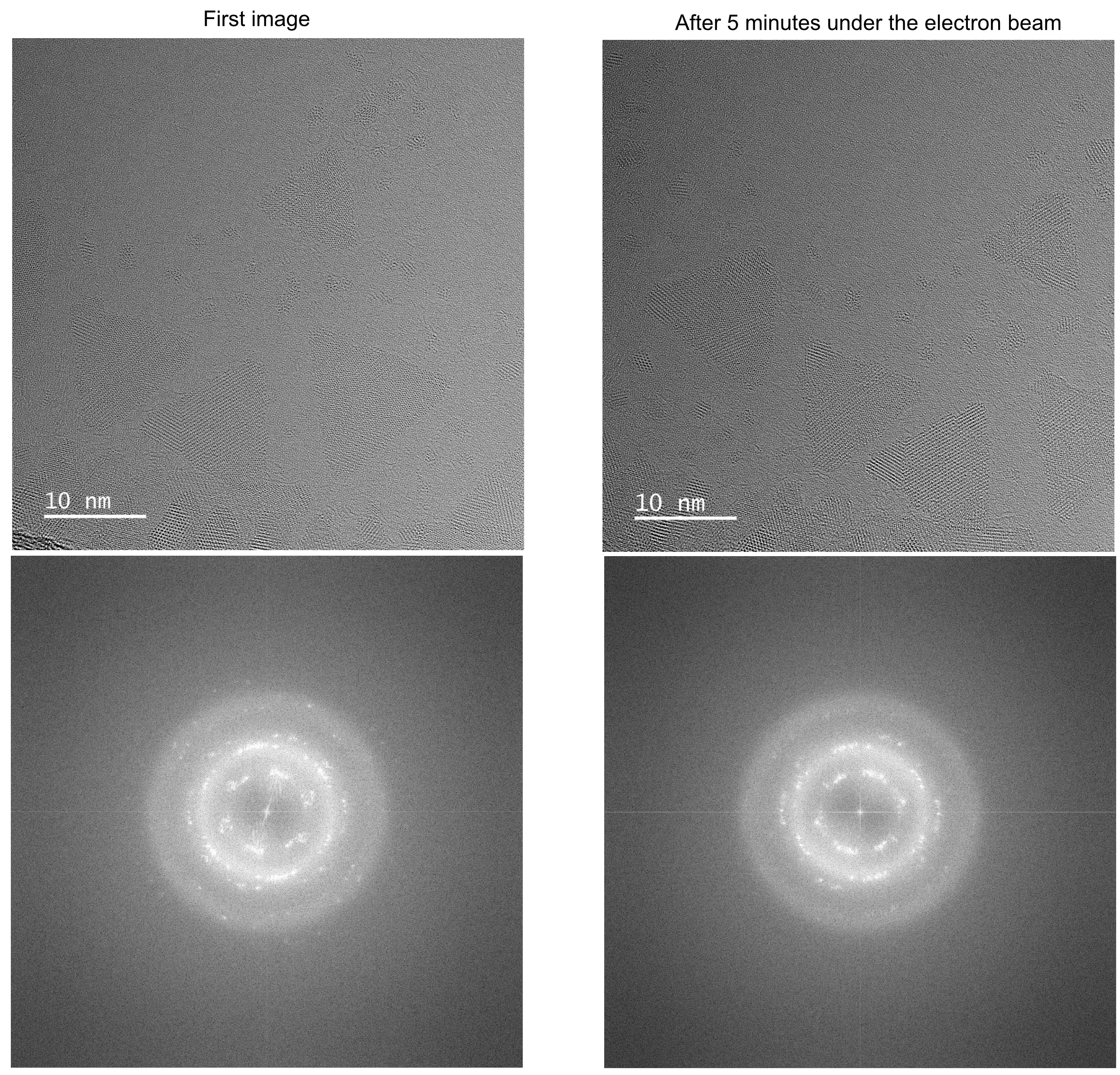}
    \caption{TEM images of triangular NPLs and their FFT showing the electron beam effect on NPL structures. We observe that the spatial information at high spatial frequency is partially lost and that the lattice parameter decreases of around 2\%}
    \label{figSI:beameffect}
\end{figure}

\begin{figure}[htbp] 
   \centering
    \includegraphics[width=0.9\linewidth]{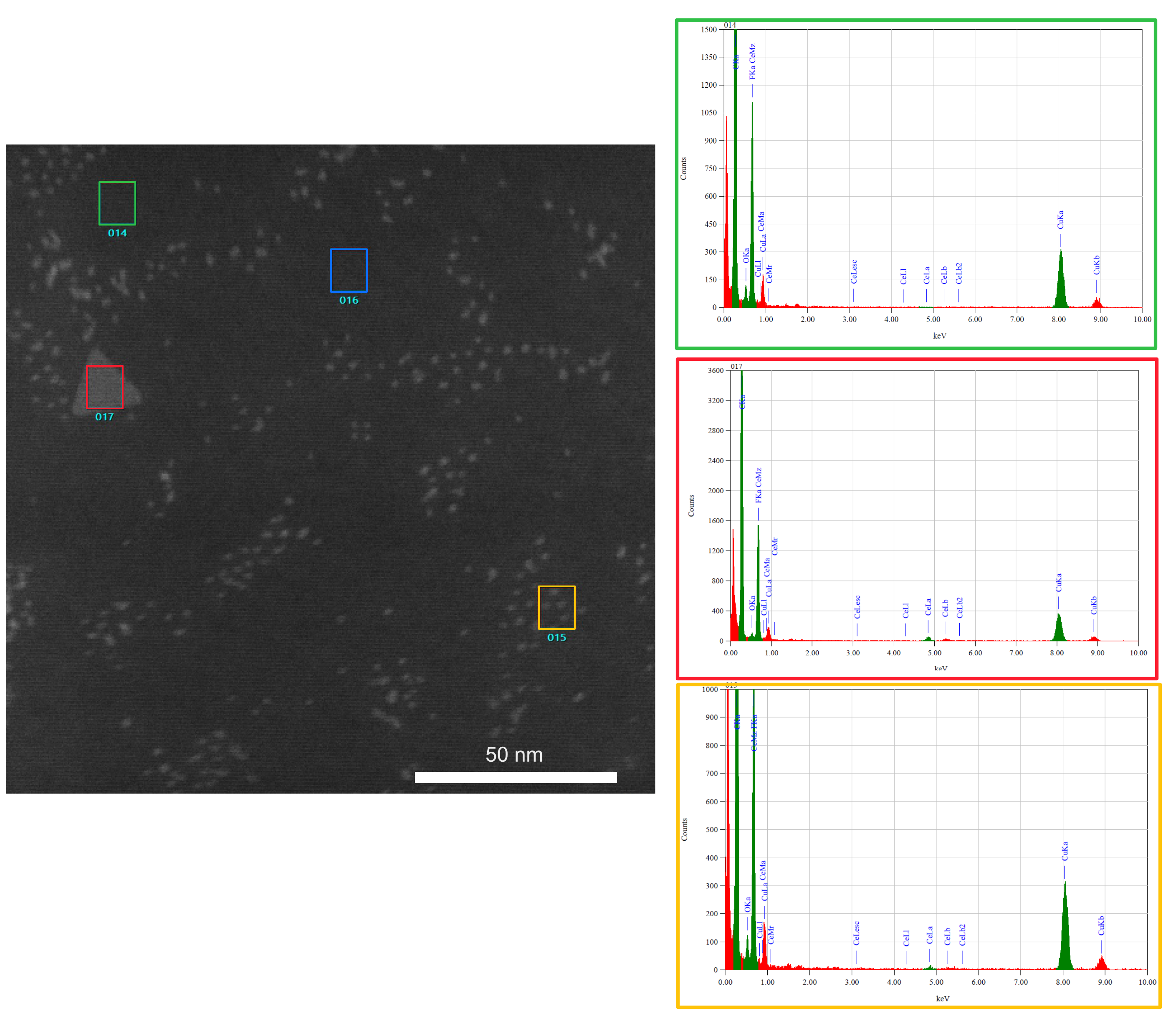}
    \caption{STEM - EDX analysis in different areas. The EDX specra on the right are framed in colors corresponding to the areas shown in the STEM image (left). Note that cerium is only detected in nanoplatelets (area framed in orange) while oxygen and fluorine are detected in all the probed areas} 
    \label{figSI:EDX}
\end{figure}

\begin{figure}[htbp] 
   \centering
    \includegraphics[width=0.9\linewidth]{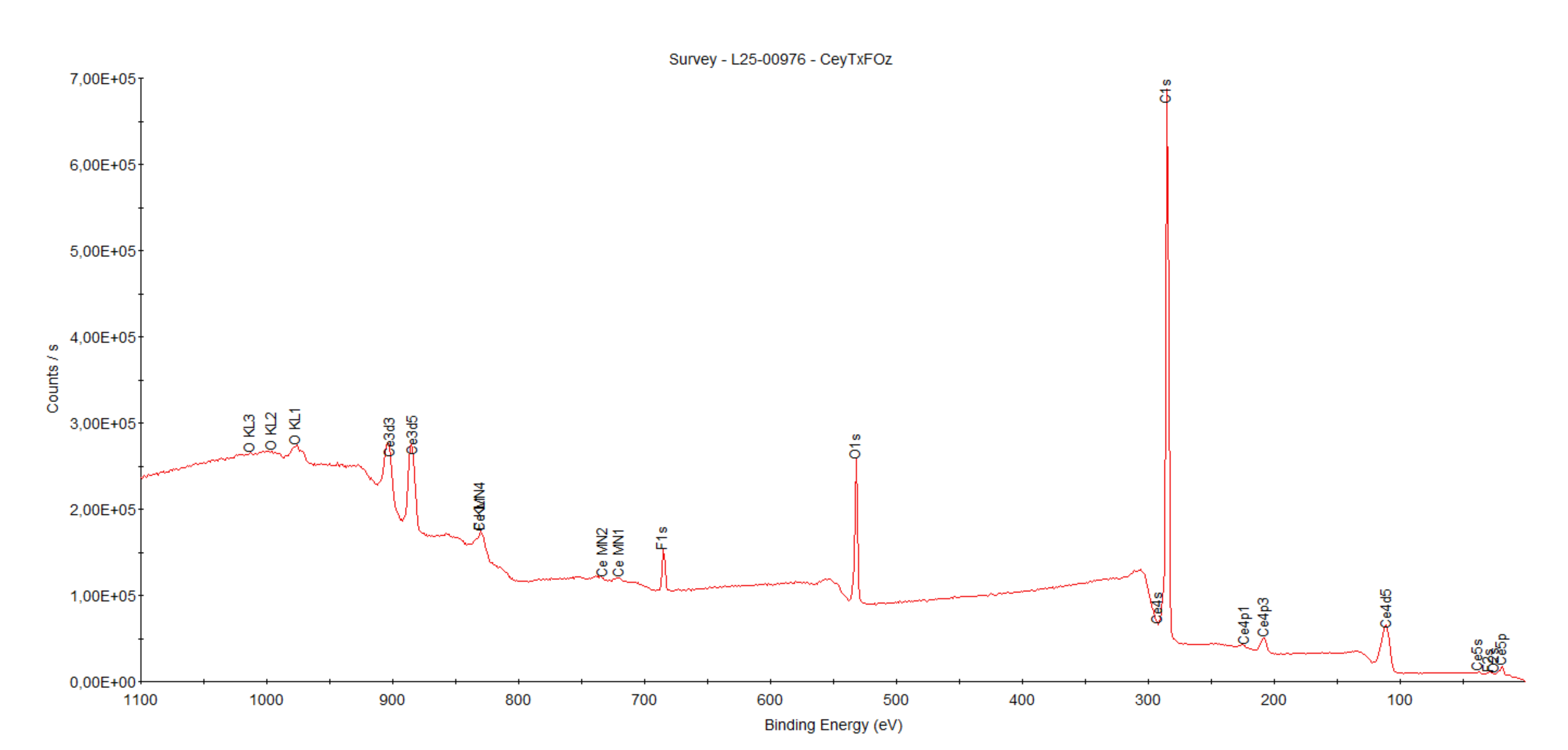}
    \caption{Survey XPS spectrum of the NPL powder showing the presence of the different elements: Ce, C, F, O.} 
    \label{figSI:XPS_survey}
\end{figure}

\begin{figure}[htbp] 
   \centering
    \includegraphics[width=0.9\linewidth]{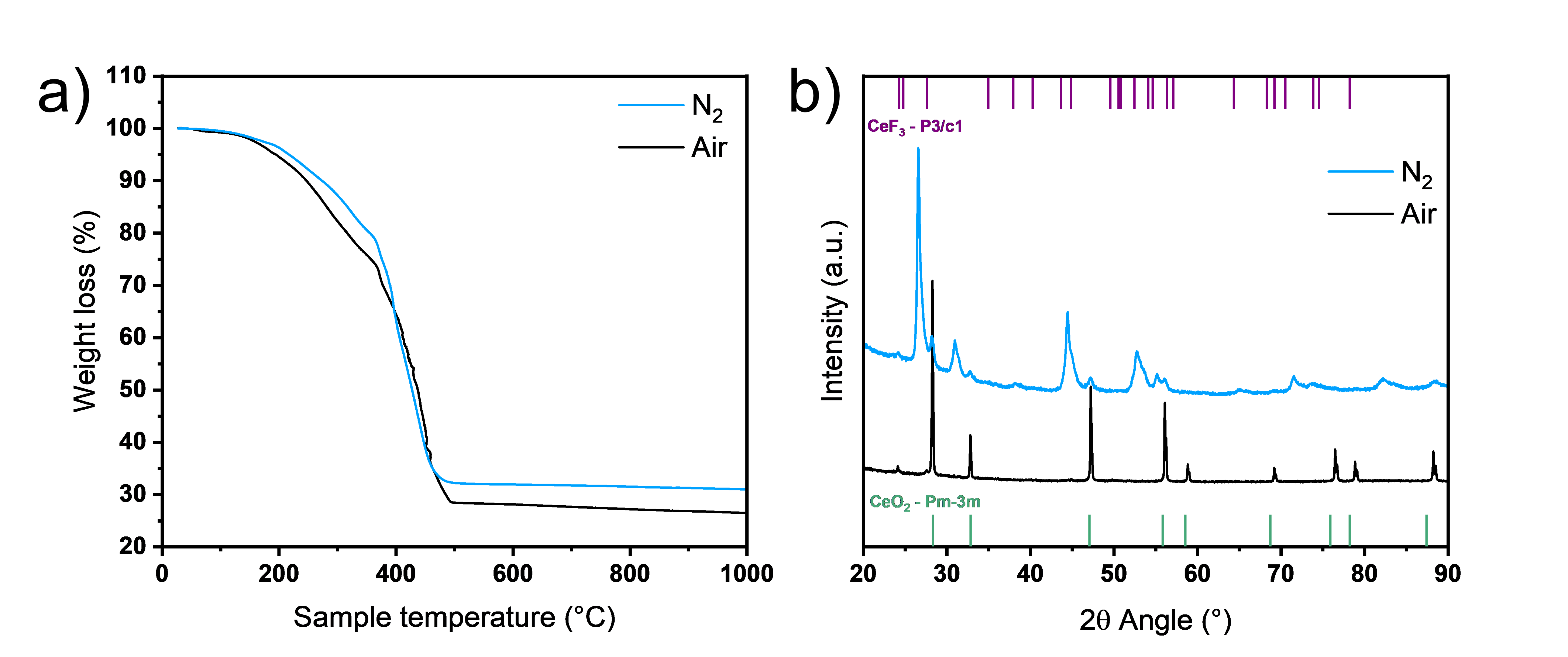}
    \caption{A) Thermogravimetric analysis of the NPLs performed in air and under \ce{N2}. B) X-ray diffraction patterns of the residues after TGA of the same NPL sample performed in air and under \ce{N2}.}
    \label{figSI:TGA}
\end{figure}

\begin{figure}[htbp] 
    \centering
    \includegraphics[width=0.8\linewidth]{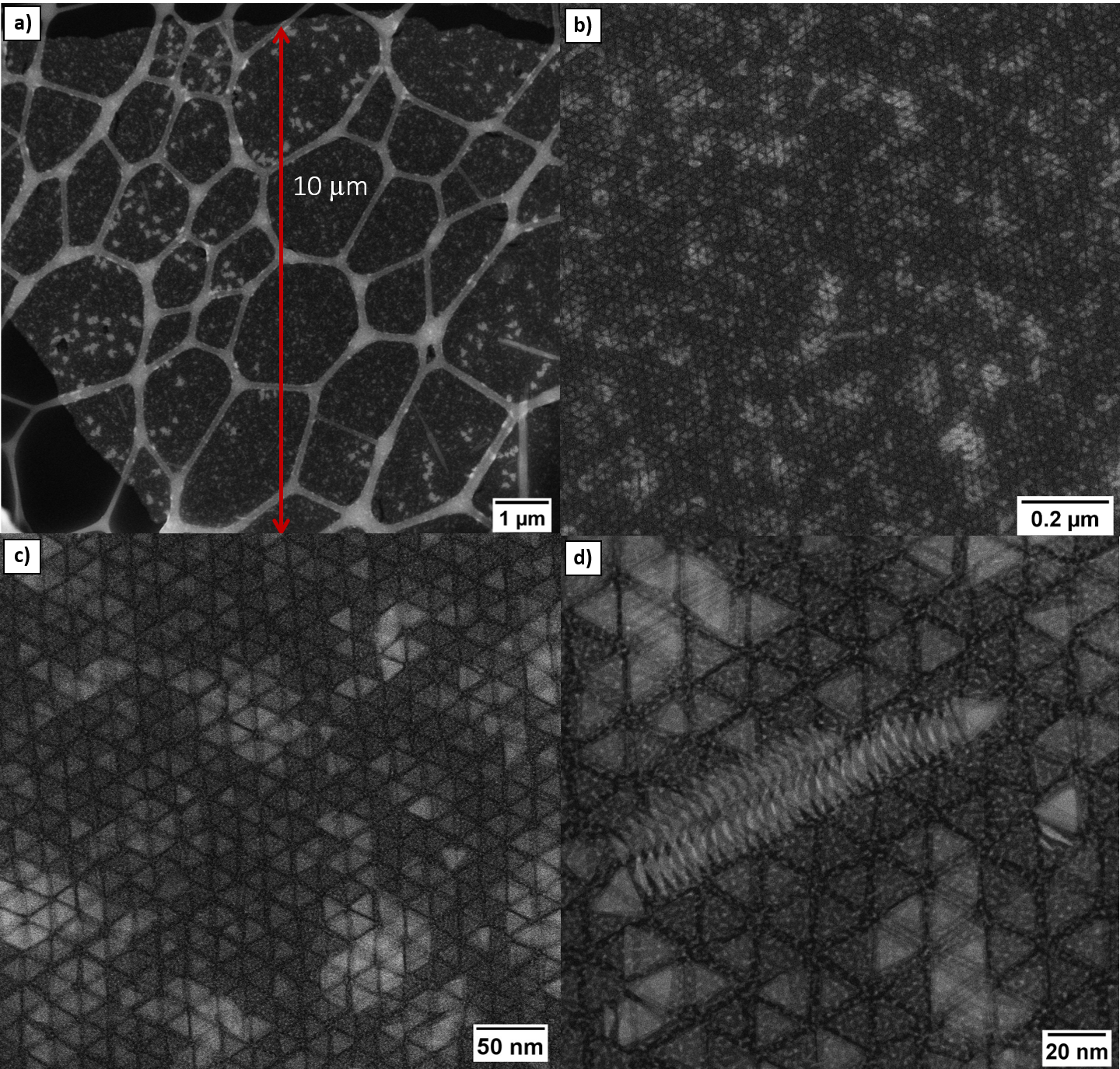}
    \caption{STEM images of triangular cerium-based NPLs after the evaporation of cyclohexane on DEG.} 
    \label{figSI:STEM-SA-cyclohexane}
\end{figure} 

\begin{figure}[htbp]
   \centering
    \includegraphics[width=0.9\linewidth]{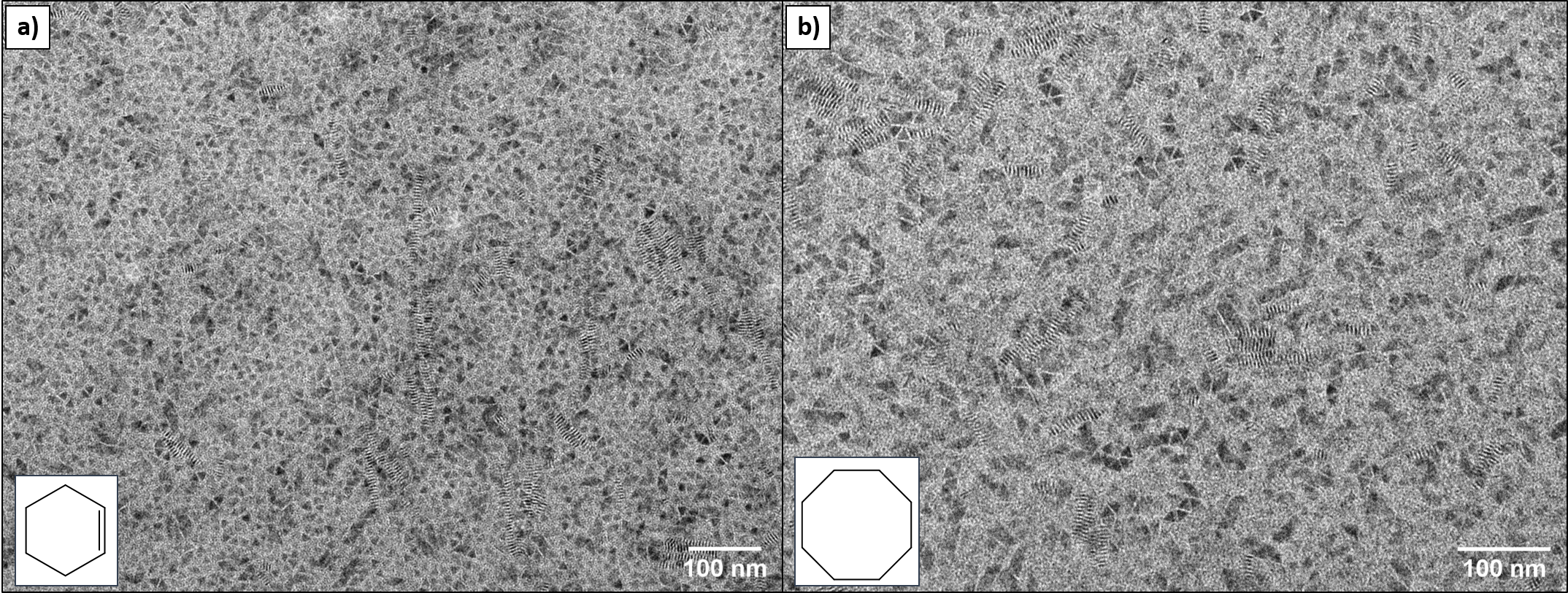}
    \caption{TEM images after the evaporation of (a) cyclohexene and of (b) cyclooctane NPL dispersion on DEG.} 
    \label{figSI:TEM-SA-cyclohexene,cyclooctane}
\end{figure} 

\begin{figure}[htbp] 
   \centering
    \includegraphics[width=0.9\linewidth]{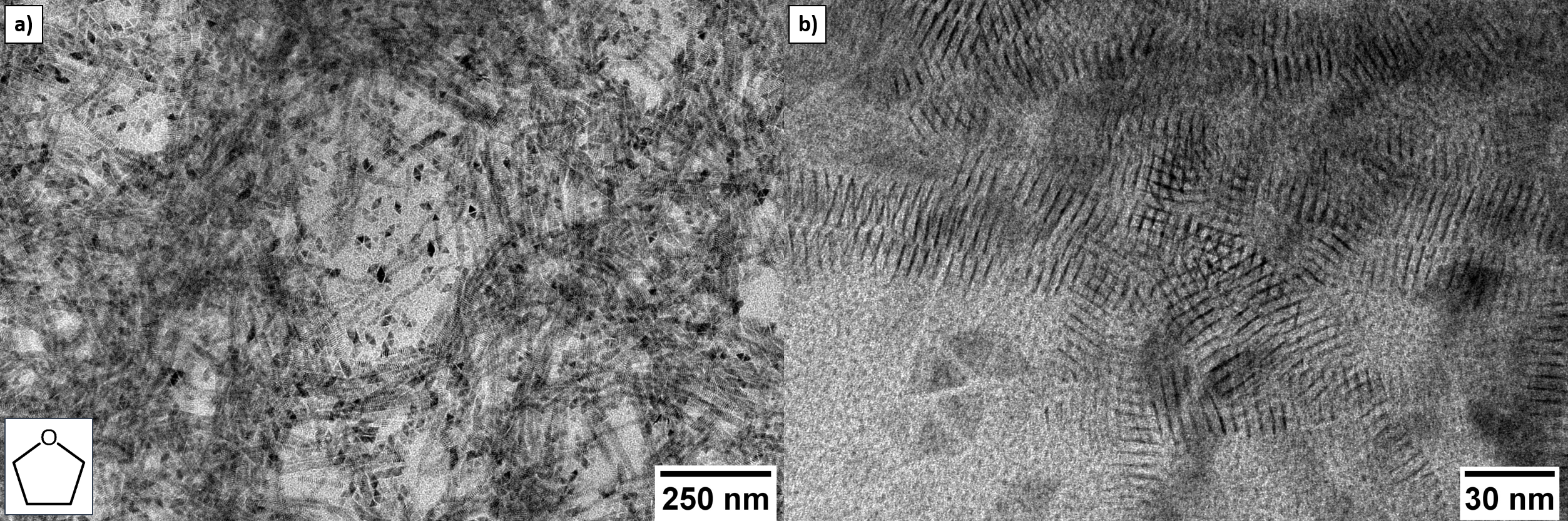}
    \caption{TEM images after the evaporation of a THF NPL dispersion on DEG.} 
    \label{figSI:TEM-SA-THF}
\end{figure} 

\begin{figure}[htbp] 
    \centering
    \includegraphics[width=0.9\linewidth]{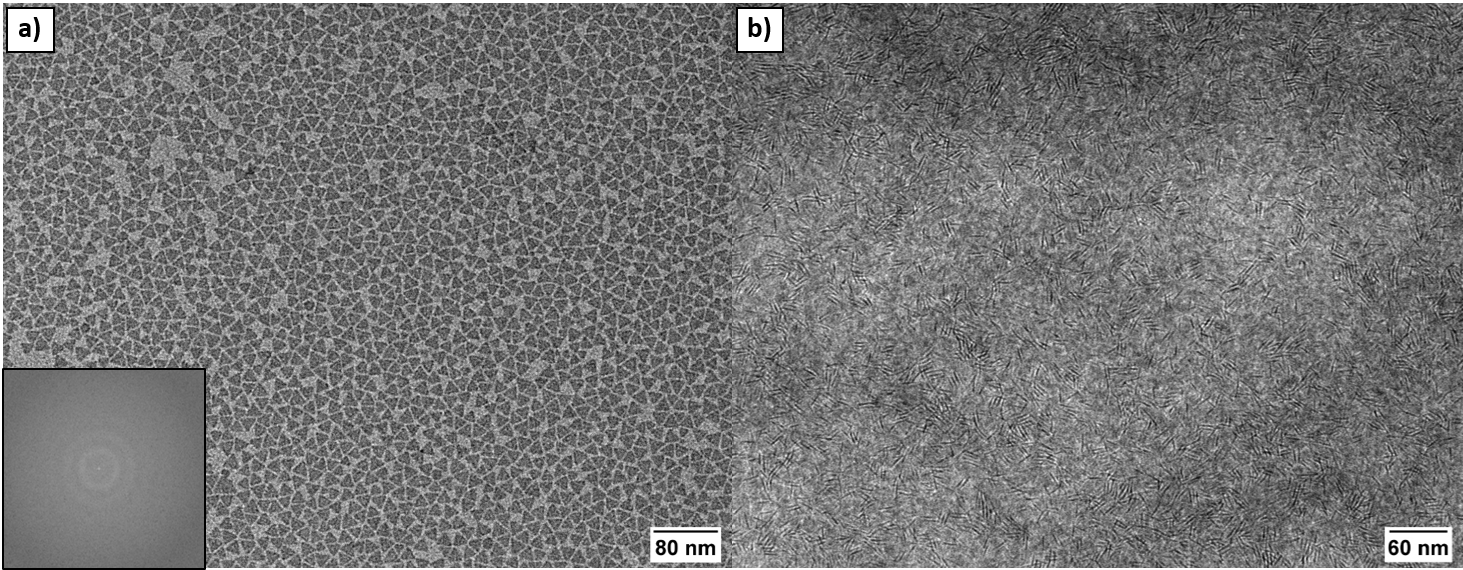}
    \caption{TEM images cerium-based NPLs after the evaporation of a hexane dispersion on DEG. a) NPLs in a plastic symmetry. b) NPLs in edge-up configurations.} 
    \label{fig:TEMCe2O3hexane}
\end{figure} 

\clearpage
\newpage

\subsection{Supplementary tables}
\begin{table}[H]
    \centering
    
    \begin{tabular}{|C{2.5cm}||C{2.5cm}|C{2.5cm}|C{2.5cm}|C{2.5cm}|C{2.5cm}|}
       \hline
        & C & O  & Ce & F & F/Ce  \\
        \hline
        sampling point 1 & 85.8 & 10.4 & 1.7 & 2.1 & 1.24 \\
       \hline
       sampling point 2 & 86.9 & 9.7  & 1.4 & 2.1 & 1.50 \\
    \hline
    \end{tabular}
    \caption{Elementary composition as measured by XPS in atomic \% for two different sampling points on the same nanoplatelet sample. The typical sensitivity is comprised between 0.1 and 0.5 at\%.}
    \label{XPS_comp}
\end{table}

\begin{table}[H]
    \centering
    
    \begin{tabular}{|C{2.5cm}|C{2.5cm}|C{2.5cm}|C{2.5cm}|C{2.5cm}|C{3.5cm}|}
       \hline
       Solvent & Relative polarity  & Vapor pressure at 20°C (hPa) & Dielectric constant  & Hildebrand parameter (MPa$^{1/2}$) & Self-assembly outcome \\
        \hline
       Toluene & 0.099 & 29 & 2.38 (25°C) & 18.3 & long range, edge up  \\
       Cyclohexane & 0.006 & 104 & 2.02 (20°C) & 16.8 & long range, face up \\
       Hexane & 0.009 & 160 & 1.88 (25°C) & 14.9 & no order, face up \\
       Tetrahydrofuran & 0.207 & 200 & 7.58 (25°C) & 18.5 & short range, edge up \\
    
    \hline
    \end{tabular}
    \caption{Main physicochemical parameters of solvents and outcome of NPL assembly at the liquid-liquid interface for toluene, cyclohexane, hexane, and THF. Values from \cite{reichardt2011}.}
    \label{tabSI:solvents}
\end{table}

\subsection{Atomic content calculation from TGA measurement}
We assume that the starting material is composed of ligands and oxyfluoride nanoplatelets with the generic formula CeO$_y$F$_z$. To simplify the calculation, we assume that we start with 100 g of material. In air, the residue is pure \ce{CeO2} (molar mass: 172.115 g/mol) and weighs 28.46 g. Hence, there is n=28.46/172.115=0.1654 moles of \ce{Ce} in the starting material. Next, we assume that the residue from decomposition in \ce{N2} is composed of $a$ moles of \ce{CeO2} and $b$ moles of \ce{CeF3} (molar mass: 197.12 g/mol). We can write $n=a+b$ since all the cerium from the two compounds comes from the initial NPL and $172.115 \times a + 197.12 \times b = 32.22$. This yields a=0.0153 and b=0.1500. Hence, the molar ratio Ce:F is 1:2.7.\\

To estimate the surface ligand density, we assume that the loss below 500$^{\circ}$C corresponds to the ligand mass. The surface density is then calculated using:
\begin{equation}
    \omega_l=\dfrac{1}{1+\dfrac{X}{\Sigma}},
\end{equation}
where $X=\dfrac{\mathcal{N}_AV_{NPL}\rho_{NPL}}{\mathcal{A}_{NPL}M_l}$, $\mathcal{N}_A$ is the Avogadro constant, $\mathcal{A}_{NPL}$, $V_{NPL}$ and $\rho_{NPL}$ are the surface area, the volume and the density of the NPL, respectively.
The volume is calculated for an edge of 17 nm and a thickness of 1.4 nm. The NPL core density is linearly interpolated between that of \ce{CeF3} (6.16 g/cm$^{3}$) and \ce{Ce2O3} (6.86 g/cm$^{3}$) with the known composition which yields 6.23 g/cm$^{3}$. With these numerical values, we find $\omega_l$=3.98 nm$^{-2}$. Since the oxygen content may have been under-evaluated, doing the same calculation with the density of \ce{Ce2O3} gives a higher bound for the surface density of 4.39 ligands per nm$^{2}$.

\end{document}